\pdfoutput=1
\documentclass[fleqn,usenatbib]{mnras}

\usepackage{newtxtext,newtxmath}

\usepackage[T1]{fontenc}

\DeclareRobustCommand{\VAN}[3]{#2}
\let\VANthebibliography\thebibliography
\def\thebibliography{\DeclareRobustCommand{\VAN}[3]{##3}\VANthebibliography}

\defcitealias{Barnes2020}{B20}

\usepackage{graphicx}	
\usepackage{amsmath}	
\usepackage{orcidlink}  
\usepackage{multirow}   
\usepackage{textgreek}  

\newcommand{\lnL}{\textrm{ln}\mathcal{L}} 

\title[DMPP-3 Revisited]{DMPP-3: confirmation of short-period S-type planet(s) in a compact eccentric binary star system, and warnings about long-period RV planet detections}

\author[A. T. Stevenson et al.]{
Adam T. Stevenson$^{1}$\thanks{E-mail: adam.stevenson@open.ac.uk}\orcidlink{0000-0003-2399-7619},
Carole A. Haswell$^{1}$\orcidlink{0000-0002-8050-1897},
John R. Barnes$^{1}$\orcidlink{0000-0001-6105-2902},
Joanna K. Barstow$^{1}$\orcidlink{0000-0003-3726-5419} and
\newauthor{~~Zachary O. B. Ross$^{1}$\orcidlink{0000-0002-3213-9643}}\\
\\
$^{1}$School of Physical Sciences, The Open University, Milton Keynes MK7 6AA, UK
}

\date{Accepted XXX. Received YYY; in original form ZZZ}

\pubyear{2023}

\begin{document}
\label{firstpage}
\pagerange{\pageref{firstpage}--\pageref{lastpage}}
\maketitle

\begin{abstract}

We present additional HARPS radial velocity observations of the highly eccentric ($e \sim 0.6$) binary system DMPP-3AB, which comprises a K0V primary and a low-mass companion at the hydrogen burning limit. The binary has a 507\,d orbital period and a 1.2\,au semi-major axis. The primary component harbours a known 2.2 M$_{\earth}$ planet, DMPP-3A\,b, with a 6.67 day orbit. New HARPS measurements constrain periastron passage for the binary orbit and add further integrity to previously derived solutions for both companion and planet orbits. \textit{Gaia} astrometry independently confirms the binary orbit, and establishes the inclination of the binary is $63.89 \pm 0.78 ^{\circ}$. We performed dynamical simulations which establish that the previously identified $\sim$800\,d RV signal cannot be attributed to an orbiting body. The additional observations, a deviation from strict periodicity, and our new analyses of activity indicators suggest the $\sim$800\,d signal is caused by stellar activity. We conclude that there may be long period planet ‘detections’ in other systems which are similar misinterpreted stellar activity artefacts. Without the unusual eccentric binary companion to the planet-hosting star we could have accepted the $\sim$800\,d signal as a probable planet. Further monitoring of DMPP-3 will reveal which signatures can be used to most efficiently identify these imposters. We also report a threshold detection (0.2 per cent FAP) of a $\sim$2.26~d periodicity in the RVs, potentially attributed to an Earth-mass S-type planet interior to DMPP-3A\,b.

\end{abstract}




\section{Introduction}
\begin{figure}
	\includegraphics[width=0.95\columnwidth]{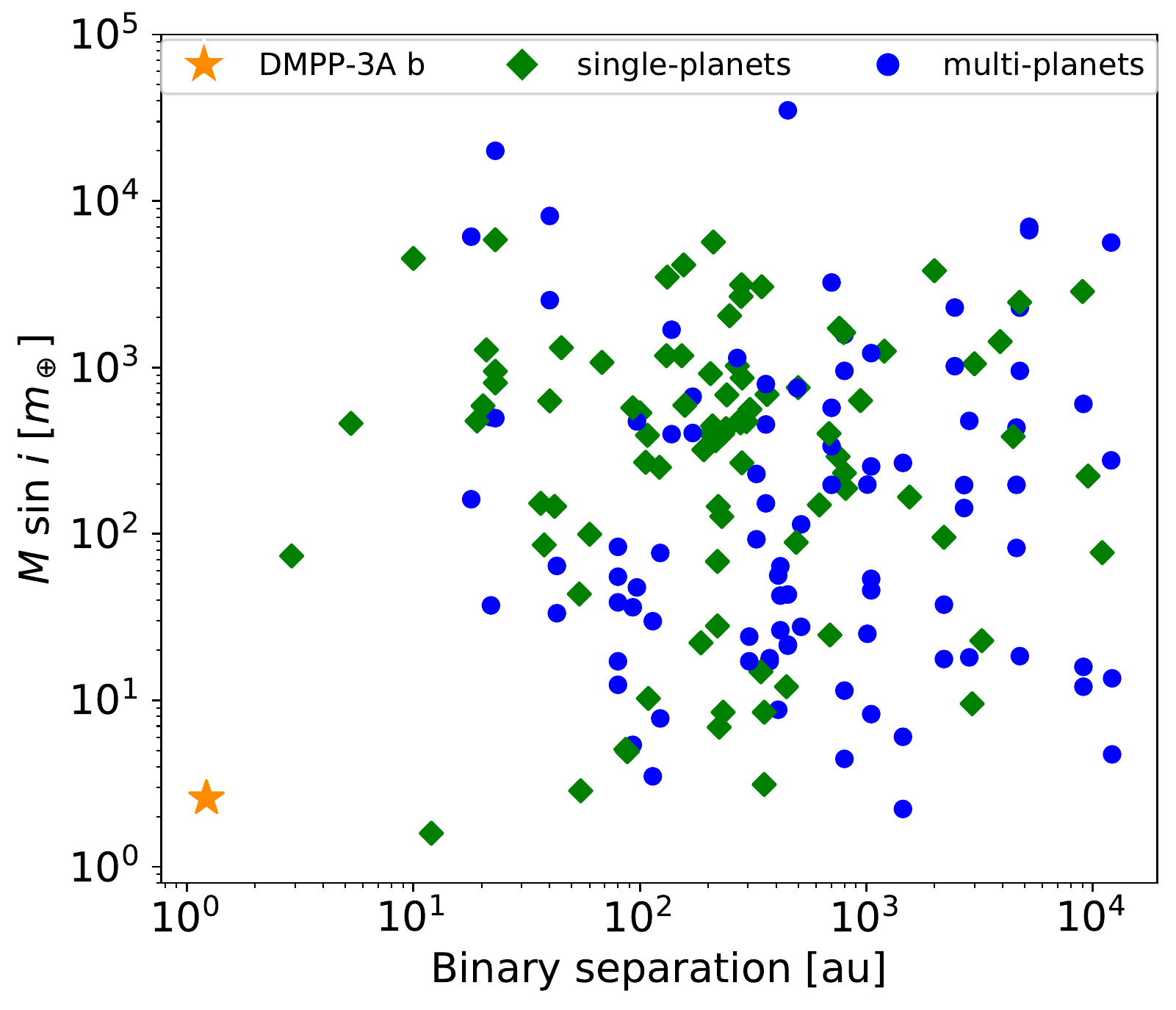}
    \caption{Planet mass as a function of binary separation for all S-type planets. Adapted from Figure 2 in \citet{2021AJ....162..272S}, we have extended the sample of stars (to include other detection methods, see \url{http://exoplanet.eu}) in order to highlight the outlying position of DMPP-3A\,b. The green diamonds correspond to single-planet systems, the blue circles denote multiple planets in a binary system, and the orange star shape identifies DMPP-3A\,b. Kepler-693A\,b resides in the next closest binary, with separation $a_{\textrm{AB}} = 2.90$~au.}
    \label{fig:my_demographics}
\end{figure}

DMPP-3 is a unique eccentric binary star system, where a hot super-Earth planet was found to orbit one of the stars in a close binary pair (\citealt{Barnes2020}, hereafter \citetalias{Barnes2020}). The circumprimary (S-type) planet orbits the primary star DMPP-3A (HD\,42936), a slowly rotating K0V star. The highly eccentric ($e=0.6$) very low mass stellar companion DMPP-3B is just above the mass required to sustain hydrogen burning ($M_{\textrm{B}}=82.5$ M$_{\textrm{jup}}$). The DMPP-3AB orbit has a semi-major axis of $a_{\textrm{AB}} = 1.23$~au. Without the context of hosting an S-type planet, DMPP-3AB is not a particularly close binary, but it is the most compact binary system to harbour an S-type planet observed thus far.
It is also one of the few systems containing a radial velocity (RV) detected S-type super-Earth around an FGK star
 (\citetalias{Barnes2020}; \citealt{2021AA...654A.104..Unger,2022AA...665A.154..Barros}). \citet{2021AJ....162..272S} compare the known S-type RV discoveries, and show that DMPP-3A\,b is an extreme outlier in their sample. In Figure~\ref{fig:my_demographics}, we have similarly plotted the demographics of all known S-type planets, i.e. we have included those discovered through their transits. DMPP-3A\,b lies in the bottom left hand corner, with lowest separation and third lowest projected planetary mass in the sample. Only two other planets in this sample have a binary separation less than 10~au, but with minimum masses more than an order of magnitude higher. DMPP-3 provides an excellent opportunity to challenge the current planetary system formation and evolution models, through studying an extreme and previously unseen system configuration.  

The scarcity of DMPP-3A b analogues found in the general demographics reported by \citet{2021AJ....162..272S} can arise from a few factors. In their work, the authors discuss the observational biases involved in RV surveys, which often disregard binary star systems. The stars in close binaries are typically unresolved, and the recorded spectrum is a blend of light from both objects. If the mass ratio between the two bodies is relatively equal, they will contribute a similar amount of light, and confuse the derived RVs \citep{2021AJ....162..272S}. For these so-called `double-lined' spectroscopic binaries, the spectra consist of two sets of superimposed lines, which require novel techniques and extensive analyses to extract the signal of each component \citep{Konacki2009, Konacki2010}. The results can still suffer from residual scatter due to the blending of the spectra.

Where the two stars in a stellar binary have very different luminosities, as is the case for DMPP-3AB, the light comes predominantly from the primary component and the spectrum effectively reveals only a single set of stellar lines. In this case there is no need for complex deconvolution to recover individual spectra. It is possible to obtain very precise radial velocities on such single-lined binaries, as demonstrated by \citet{Standing2022} and \citet{Triaud2022}. These studies focus on binary signal subtraction to detect circumbinary (P-type) planets orbiting outside the inner stellar pair, and are able to reach a residual root mean squared scatter of 3~m\,s$^{-1}$ \citep{Standing2022}. This is a different system architecture to the circumprimary (S-type) planet hosts shown in Fig.~\ref{fig:my_demographics} and under discussion in the present paper.

The double-lined nature of many binary observations therefore hinders the detection of low mass planets, almost certainly causing them to be under-represented in the samples. However, some of the scarcity of known DMPP-3 analogues must be due to the influence that multiplicity has on the formation and retention of these planets \citep{Holman1999, Jang-Condell2015, Kraus2016, Marzari2019}. A highly eccentric, close in companion dramatically affects the dynamical stability of any other orbiting bodies, and would truncate the protoplanetary disc through dynamical perturbations, limiting the available mass for the creation of planets.

Planetary stability in binary systems has been studied over the years in an attempt to solve the three-body problem and find regions where planets can reside such systems \citep{Dvorak1986,Mardling1999}. \citet{Holman1999} performed numerical simulations to investigate the possible stable S-type orbits for companions with a variety of mass ratios, eccentricities, and semi-major axes. They developed a semi-empirical formula to determine the critical semi-major axis: a threshold value exterior to which a planet cannot orbit. DMPP-3A b is within their `safe' zone (which extends out to 0.16~au from the primary star in this system), a finding confirmed through simulations in \citetalias{Barnes2020}. 

Whilst DMPP-3A\,b's present orbit is stable, we must also consider challenges to the formation of this system. The protoplanetary disc would be truncated by the presence of another massive body, an effect strongest for large eccentricity and small semi-major axis of companion orbit: two features this system exemplifies \citep{Jang-Condell2015}. If S-type planets do begin to form, such systems can be short-lived, with secular orbital evolution resulting in ejection on timescales of millions to billions of years \citep{Kraus2016}. DMPP-3A is 9.6~Gyr old (Table~\ref{tab:dmpp-3aprops}). It seems likely the architecture of the DMPP-3 system has evolved to the current configuration through dynamical interactions. 

This system was selected for intensive high precision, high cadence RV observations due to anomalously low chromospheric emission, characterised by the  \hbox{$\log R^{\prime}_{\rm HK}$} metric, and attributed to the presence of circumstellar material around the primary star. This is hypothesized to be supplied by the loss of material from close-in planets  \citep{Haswell2020}, and is therefore indicative of short period planets. The most dramatic examples of mass-losing short period planets are the so-called catastrophically disintegrating exoplanets (CDEs). These are rocky planets with very short orbital periods. They are heated so intensely that the rocky surface is vaporised and carried off in a thermal wind.
The DMPP-3 system seems likely to be in a short-lived planetary mass-losing phase, perhaps following a dynamical reconfiguration. Employing an approximate relation for planetary temperature to orbital distance \citep{2012ApJ...752....1Rappaport}, a CDE surface temperature of approximately 2000~K \citep{Jones2020} would correspond to a $\sim$1.6~d orbit around DMPP-3A. Small planets on 1--2 day orbits are therefore likely to lose mass.

Despite the often disastrous influence close-in eccentric binary companions are understood to have on planet formation (exemplified by the dearth of planets found in binaries with separation $\lesssim10$~au seen in Fig.~\ref{fig:my_demographics}), the occurrence rate for hot Jupiter planet hosts having a stellar companion is twice as high as the binarity ratio of field stars with projected separations 20 -- 10000 au \citep{Cadman2022}. A secondary star can modify the projected spin-orbit angle between primary star and planet (observed through measuring the Rossiter--McLaughlin effect), exciting secular evolution and interactions that cause a departure from coplanarity for all orbiting bodies \citep{2018MNRAS.479.1297Martin,Franchini2020, 2021MNRAS.507.3593..Moe,Best2022}. Recent evidence indeed suggests that orbital misalignment between a planet and a secondary stellar companion is relatively common for close-in giants ($P_\textrm{P} <10$~d, $R_\textrm{P} > 4~R_{\earth}$), with systems tending to favour a polar orbit. In comparison, smaller  ($a_\textrm{P} <1$~au, $R_\textrm{P} < 4~R_{\earth}$) planets are mostly found in less inclined orientations relative to a companion star, with inclination angles between the planetary and companion orbits of $\sim$10--50 degrees \citep{Behmard2022}.

The architecture we see in DMPP-3 could be created through the eccentric Kozai--Lidov mechanism (EKL: \citealt{1962..Lidov}; \citealt{1962..Kozai}). In hierarchical triple systems where the inclinations are not co-planar, a perturbing companion can stir up oscillations in the inclination and eccentricity, caused by the overall requirement for angular momentum to be conserved \citep{2016ARA&A..54..441Noaz}. This is thought to be the cause of non-zero eccentricities of S-type planets in binaries. Although close-in planets have short tidal circularisation timescales, gravitational interactions with a binary companion can excite eccentricity on an even shorter dynamical timescale.

The EKL may have driven the migration for some hot Jupiter planets \citep{2015Icar..248...89Rubie,2015Icar..258..418Morbidelli,2022AJ....163..227Angelo..kepler1656b}. A companion can perturb the planet orbit into gradually more eccentric configurations over secular timescales (far longer than the orbital periods involved in the system), and with increased eccentricity the periastron will gradually move closer to the primary star. Tidal forces on the planet and star then tend to shrink the orbit and circularise it, creating a close-in planet \citep{2007ApJ...669.1298F, 2016ARA&A..54..441Noaz}. 

Dynamical interactions can significantly alter orbital parameters of the system, but can also change the structure altogether. \citet{2018MNRAS...Gong} highlight that whilst formation of S-type planets in a close binary is challenging, the formation of circumbinary (P-type) planets would be comparatively easier to achieve - and should be relatively common throughout the Universe. So far, a small number of transiting P-type planets have been discovered, by the \textit{Kepler} \citep{KeplerMission} and \textit{TESS} \citep{TessMission} missions (eg. TIC 172900988\,b; \citealt{Kostov2021}). The first P-type planet discovered purely through RVs, TOI-1338/BEBOP-1\,c, was also recently announced by \citet{Standing2023}. Despite these discoveries, no planets have yet been found with orbits of the scale of $\sim3$~au around a $\sim1$~au binary.

To study the evolution of such systems, \citet{2018MNRAS...Gong} explore planet-planet scattering,  occurring for multiple planets when the eccentricities are stirred up (perhaps by the close binary pair developing eccentricity too). Tidal interactions could cause the P-type planets to be captured by the primary star in the close binary, becoming S-type. The capture probability is low (and related to mass ratio and eccentricity of the binary), but scattering through multiple planet interactions would increase this probability by reducing the energy of the planet \citep{2018MNRAS...Gong}. The capture scenario should be possible for close binaries with semi-major axes of $0.5-3$~au, such as the DMPP-3 system. DMPP-3 is therefore useful for investigating planetary system  evolution. Several physical processes with important inferred roles in sculpting the demographics of the Galaxy's short period planets could be strongly at play in creating and maintaining the exotic architecture we observe.

This paper presents additional  RV data for DMPP-3. The new observations are described in Section~\ref{sec:newobs}; updated stellar parameters of DMPP-3A are discussed in Section~\ref{sec:dmpp-3aparams}; the RV analysis, refined system parameters, and further periodicities identified are described in Section~\ref{sec:rvanal}. We discuss mutually inclined orbital simulations in Section~\ref{sec:3Dsims}, reporting the resulting dynamical timescales for system disruption. In Section~\ref{sec:activity} we consider the issue of stellar activity and how that impacts the periodic variability of the spectra. Our results are then discussed in Section~\ref{sec:discussion}, and we summarise the findings and conclude in Section~\ref{sec:conclusions}.

\begin{table}
	\centering
	\caption{HD42936 (DMPP-3A) stellar parameters. Updated from \citetalias{Barnes2020}, the values are shown with corresponding 1$\sigma$ uncertainties where appropriate, and references are given for each external source. The remaining parameters were re-calculated for this work with the latest version of the \textsc{species} code \citep{2018..SPECIES.I}, using the ten highest S/N spectra from the DMPP HD42936 dataset. SIMBAD data are accessed from~\url{http://simbad.u-strasbg.fr}.}
	\label{tab:dmpp-3aprops}
	\begin{tabular}{lccr} 
		\hline
		Parameter & Value & Reference\\
		\hline
		Spectral type & K0V & \citet{1975mcts.book.....Houk} \\
		Parallax (mas) & $21.25 \pm 0.11$ & \citet{GaiaDR3paper} \\ 
		Distance (pc) & $47.06 \pm 0.25$ & \citet{GaiaDR3paper} \\ 
		$V$ & $9.09$ & SIMBAD \\ 
		$B-V$ & $0.91$ & SIMBAD \\ 
		$\log R^{\prime}_{\rm HK}$ & $-5.14 \pm 0.05$ & \citet{2011AA...531A...8Jenkins} \\
		$T_{\textrm{eff}}$ (K) & $5201 \pm 20$ &  \\
		\text{[Fe/H]} & $0.147 \pm 0.013$ &  \\
		$\log g$ (cm s$^{-2}$) & $4.266 \pm 0.045$ & \\
		$v\sin{i}$ (km s$^{-1}$) & $3.17 \pm 0.1$ & \\
		$v_{\textrm{mac}}$ (km s$^{-1}$) & $1.58 \pm 0.10$ & \\
		$R_{*} (\textrm{R}_{\sun})$ & $0.861 \pm 0.005 $ &  \\
		$M_{*} (\textrm{M}_{\sun})$ & $0.900 \pm 0.009 $ &  \\
		$L_{*} (\textrm{L}_{\sun})$ & $0.510 \pm 0.003$ & \citet{GaiaDR2.2018} \\
		Age (Gyr) & $9.6 \pm 0.8$ &  \\
		\hline
	\end{tabular}
\end{table}

\section{Observations}\label{sec:newobs}

Previous RV observations with the High Accuracy Radial Velocity Planet Searcher ({\sc HARPS}, \citealt{Harps}) enabled the binary orbit of DMPP-3AB to be identified and characterised for the first time \citepalias{Barnes2020}. The best solution indicated a highly eccentric, low mass ratio binary, with large RV excursion around periastron. Since the {\sc HARPS} observations did not sample orbital phases close to periastron, the RVs were supplemented with two further measurements from the CORALIE spectrograph at the 1.2-metre Leonhard Euler Telescope (see Fig.~\ref{fig:RVplots}, and Figure 1 in \citetalias{Barnes2020}). The CORALIE observations confirmed the RV trend predicted by {\sc HARPS}, but are an order of magnitude less precise, with RV uncertainties of 9~m~s$^{-1}$. Moreover, the use of a different spectrograph meant that the observations had to be treated as a separate dataset. This necessitated an additional RV offset parameter meaning that the two CORALIE observations do not place a strong constraint on the orbital solution. We thus secured five new observations in service mode with {\sc HARPS} around the September 2021 periastron, see Table~\ref{tab:newobs}.

\begin{table}
	\centering
	\caption{The DMPP-3 observations taken in 2021. These science observations were performed with HARPS Echelle observing technique, for consistency with the previous data. Each exposure lasted 900~s.}
	\label{tab:newobs}
	\begin{tabular}{lcccc} 
		\hline
		Date & BJD $-2450000$ & Local time (La Silla) \\
		\hline
		2021 Sept 07  & 9464.875 & 03:16:42 \\
		2021 Sept 13 & 9470.885 & 03:53:47 \\
		2021 Sept 16 & 9473.888 & 04:09:38 \\
		2021 Sept 20 & 9477.861 & 03:46:53 \\
		2021 Sept 22 & 9479.820 & 02:56:30 \\
		\hline
	\end{tabular}
\end{table}

We use a total of 106 spectra.  The five new observations we refer to as S21. The remaining observations comprise the data used by \citetalias{Barnes2020} and are as follows: eight {\sc HARPS} observations made between 2008 and 2013 by the Calan--Hertfordshire Extrasolar Planet Search ({\sc CHEPS}, \citealt{2009MNRAS.398..911J..Calan}); the main body of ninety-one {\sc HARPS} observations made between 2015 and 2018 ({\sc DMPP}); and two CORALIE observations made in 2017.

Data from the {\sc HARPS} spectrograph were reduced using the \textsc{harps-terra} software \citep{HarpsTerra} to determine the radial velocities from the wavelength-calibrated spectra through a template matching process. Ancillary measurements of activity from the cross-correlation functions (CCF) are obtained through the standard {\sc HARPS} data reduction software ({\sc drs}\footnote{\url{http://www.eso.org/sci/facilities/lasilla/instruments/harps/doc/DRS.pdf}}) pipeline. For more information on the reduction process used, see \citetalias{Barnes2020}.

\section{Stellar parameters of DMPP-3A}\label{sec:dmpp-3aparams}
Properties of DMPP-3A were derived from the stellar spectrum using the ten highest S/N spectra from the DMPP dataset with the latest version of the \textsc{species} code \citep{2018..SPECIES.I}. Along with values from the literature, these are given in Table~\ref{tab:dmpp-3aprops}. The stellar parameters which are consistent with those reported in \citetalias{Barnes2020} comprise: the $V$ and $B-V$ magnitudes; the chromospheric activity index $\log R^{\prime}_{\rm HK}$ (derived from core emission in the \ion{Calcium}{II} H\&K lines); effective temperature $T_{\textrm{eff}}$; metalicity [Fe/H]; surface gravity $\log g$; macroturbulence velocity $v_{\textrm{mac}}$;  stellar mass $M_{*}$; and age. The luminosity $L_{*}$ was not included. 

The parameters that are not consistent with values and error bounds from \citetalias{Barnes2020} are similar enough to not drastically impact the analysis. The parallax and distance are altered solely due to the availability of another \textit{Gaia} data release between the analyses, but this has no impact on any numerical solutions. 

The two remaining parameters both come from the \textsc{species} analysis using a program version updated from than that used by \citetalias{Barnes2020}. The projected rotation velocity $v\sin{i}$ changes considerably, from 1.97 to 3.17 km s$^{-1}$. Despite this increase, DMPP-3A remains an old, slowly rotating star. We will discuss the observed $v\sin{i}$ and indicators of the rotation period in Section~\ref{sec:rotation}. The stellar radius $R_{*}$ is reduced by 0.05~R$_{\sun}$ (with error bounds in \citetalias{Barnes2020} being $\pm 0.02$~R$_{\sun}$).

\renewcommand{\arraystretch}{1.5}
\begin{table*}
	\centering
	\caption{Maximum a posteriori parameters for a model with four simultaneously fitted Keplerian signals. The false alarm probabilities (FAPs) describe the confidence in the detected signals. Values are shown with uncertainties from the 68.3 per cent confidence intervals obtained from MCMC sampling. The change in BIC and $\lnL$ as signals are added is listed, providing sufficient evidence for the inclusion of additional Keplerians. The $\gamma$ values are the instrumental and reduction zero point offsets for each dataset, and the $\sigma$ values are the fitted stellar RV jitter parameters, which are shared between the four simultaneously fitted signals. Reduced chi-squared $\chi^{2}_{\textrm{r}}$ and r.m.s for the fit are also shown in the table. The subscript 0 for mean anomaly and longitude of periastron, $M$ and $\omega$, refers to these being given at reference epoch $t_{\textrm{0}}$. The columns shown for Signals 3 and 4 are Keplerian fits with eccentricities fixed at 0. Including eccentricity as a free parameter yielded no significant improvement in the BIC for either signal ($\Delta$BIC $<2$), so these signals are assumed to be sinusoidal.}
	\label{tab:objects_params}
	\begin{tabular}{lllll} 
		\hline
		Parameter & Signal 1 & Signal 2 & Signal 3 & Signal 4 \\
		~ & DMPP-3B & DMPP-3A b & ~~~~---~~~~ & DMPP-3A c \\         
	    \hline
		FAP (GLS) & $4.5\times 10^{-15}$ & $8.7\times 10^{-7}$ & $8.2\times 10^{-6}$  & $1.9\times 10^{-3}$\\
		$\Delta~\lnL$ & $\sim10^{7}$& $20.10$ & $31.48$ & $10.38$ \\
		$\chi^{2}_{\textrm{r}}$ & 1.19 & 1.26 & 1.02 & 0.95 \\
		r.m.s (m s$^{-1}$) & 1.58 & 1.34 & 0.98 &  0.97\\
		$\Delta$BIC & $\sim10^{8}$ & $16.89$ & $44.31$ &  $2.14$ \\
		$P$ (d) & 506.89$_{-0.01}^{+0.01}$ & 6.67$_{-0.01}^{+0.03}$ & 809.38$_{-0.34}^{+0.20}$ &  2.26$_{-0.10}^{+0.20}$\\
		$K$ (m s$^{-1}$) & 2657.31$_{-0.02}^{+0.33}$ & 0.82$_{-0.07}^{+0.20}$ & 3.52$_{-0.34}^{+0.20}$ &  0.52$_{-0.14}^{+0.09}$ \\
		$M_{\textrm{0}}~(\degr)$ & 126.08$_{-0.05}^{+0.03}$ & 210.26$_{-0.47}^{+0.05}$ & 196.50$_{-0.40}^{+0.02}$  &  33.50$_{-0.11}^{+0.49}$\\
		$e$ & 0.596$_{-0.001}^{+0.001}$ & 0.174$_{-0.084}^{+0.032}$ & [0, fixed] &  [0, fixed]\\
		$\omega_{\textrm{0}}~(\degr)$ & 158.88$_{-0.01}^{+0.03}$ & 52.63$_{-0.46}^{+0.10}$ & 286.71$_{-0.11}^{+0.28}$ &  47.40$_{-0.22}^{+0.08}$\\
		$M_{\textrm{p}}\sin{i}$ & $82.52\substack{+ 0.53 \\ -0.53}$~M$_{\textrm{jup}}$ & $2.22\substack{+0.50 \\ -0.28}$~M$_{\earth}$& $0.156\substack{+ 0.007 \\ -0.007}$~M$_{\textrm{jup}}$ &  $1.065\substack{+0.173 \\ -0.259}~$M$_{\earth}$\\
		$a_{\textrm{p}}$ (au) & $1.139\substack{+0.004 \\ -0.004}$ & $0.0670\substack{+0.0003 \\ -0.0002}$ & $1.641\substack{+0.006 \\ -0.005}$ &  $0.033\substack{+0.002 \\ -0.0001}$\\
		$\gamma_{\textrm{CHEPS}}$ (m s$^{-1}$) & \multicolumn{4}{c}{-669.16$_{-0.37}^{+0.37}$}\\
		$\gamma_{\textrm{CORALIE}}$ (m s$^{-1}$) & \multicolumn{4}{c}{3583.29$_{-0.41}^{+0.41}$}\\
		$\gamma_{\textrm{DMPP}}$ (m s$^{-1}$) & \multicolumn{4}{c}{-661.66$_{-0.28}^{+0.28}$}\\
		$\gamma_{\textrm{S21}}$ (m s$^{-1}$) & \multicolumn{4}{c}{-680.00$_{-0.30}^{+0.30}$}\\
		$\sigma_{\textrm{CHEPS}}$ (m s$^{-1}$) & \multicolumn{4}{c}{0.25$_{-0.11}^{+0.23}$}\\
		$\sigma_{\textrm{CORALIE}}$ (m s$^{-1}$) & \multicolumn{4}{c}{0.17$_{-0.09}^{+0.20}$}\\
		$\sigma_{\textrm{DMPP}}$ (m s$^{-1}$) & \multicolumn{4}{c}{0.11$_{-0.10}^{+0.42}$}\\
		$\sigma_{\textrm{S21}}$ (m s$^{-1}$) & \multicolumn{4}{c}{0.27$_{-0.01}^{+0.31}$}\\
		$N_{\textrm{obs}}$ & \multicolumn{4}{c}{8+2+91+5}\\
		Baseline (d/yr) & \multicolumn{4}{c}{4900 / 13.4}\\
		$t_{\textrm{0}}$ (BJD) & \multicolumn{4}{c}{2454579.56}\\
		\hline
	\end{tabular}
\end{table*}

\section{RV Analysis}\label{sec:rvanal}

For the analysis of RV data we have made use of the \textsc{exo-striker} software. This is a "transit and radial velocity interactive fitting tool for orbital analysis and N-body simulations" \citep{exostriker}. The tool takes transit and / or RV data as input, and provides quick access tabs on a graphical user interface (GUI) to a suite of useful functions. The tool can be used to fit Keplerian orbits to reflex RV signals present in the data, with periods being identified from the generalised Lomb-Scargle periodogram (GLS: \citealt{GLSZechmeister}). Whilst fitting RV solutions, the routines can simultaneously determine offsets between different datasets for the same object, which is particularly useful here as the baseline covers instrument fibre upgrades and COVID-19 shutdown. The maximum likelihood parameters for the system can be explored using the built-in \textsc{emcee} MCMC sampling functionality \citep{emcee}.

\begin{figure*}
    \includegraphics[width=0.55\linewidth]{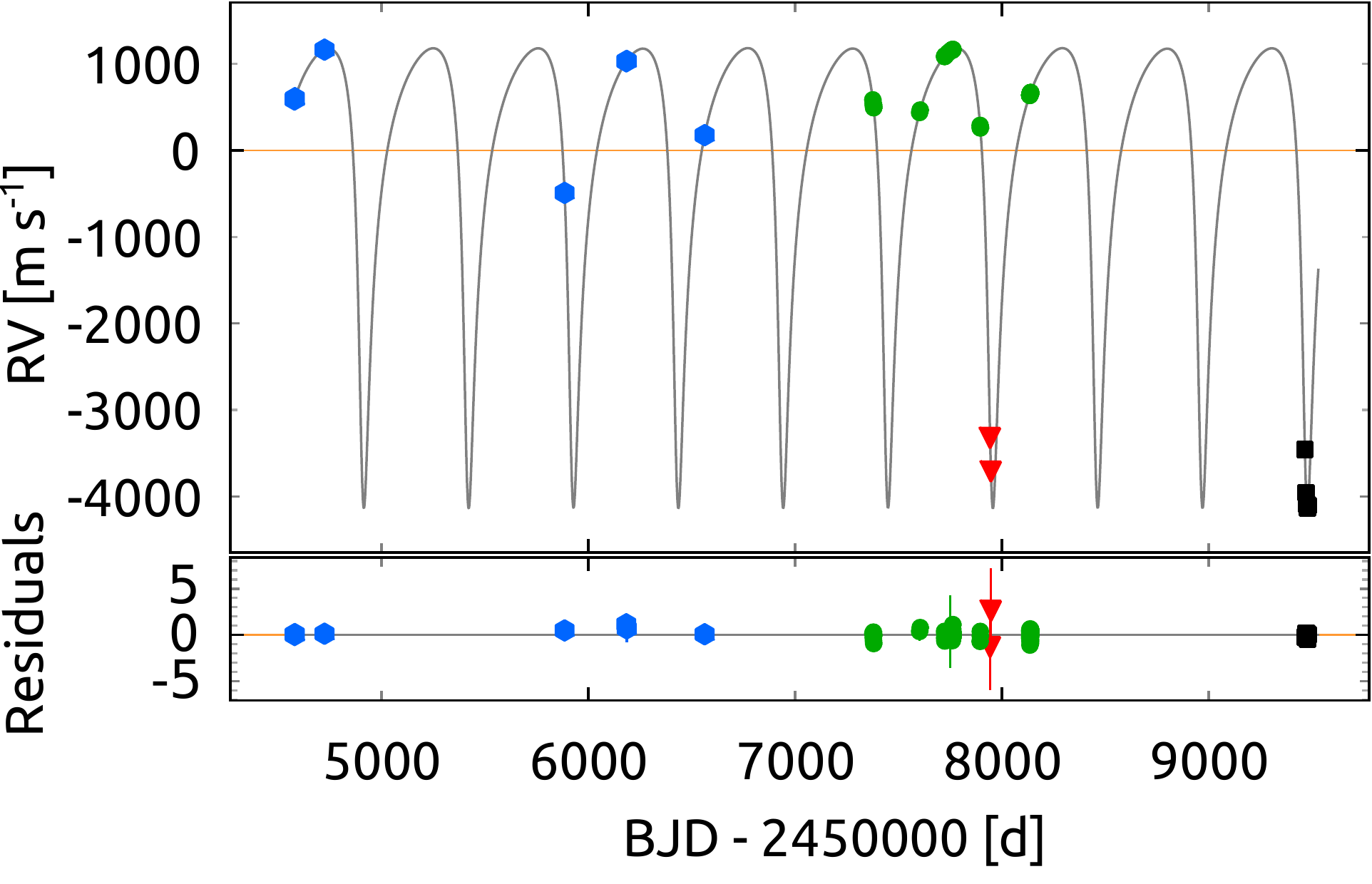}
    \includegraphics[width=0.48\linewidth]{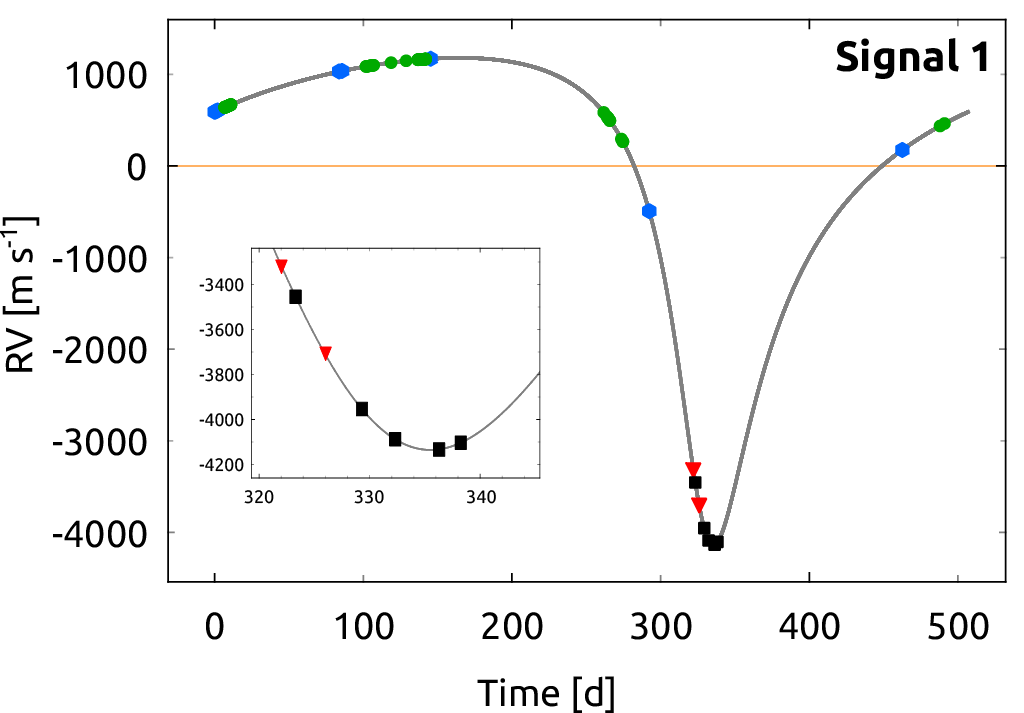}\includegraphics[width=0.48\linewidth]{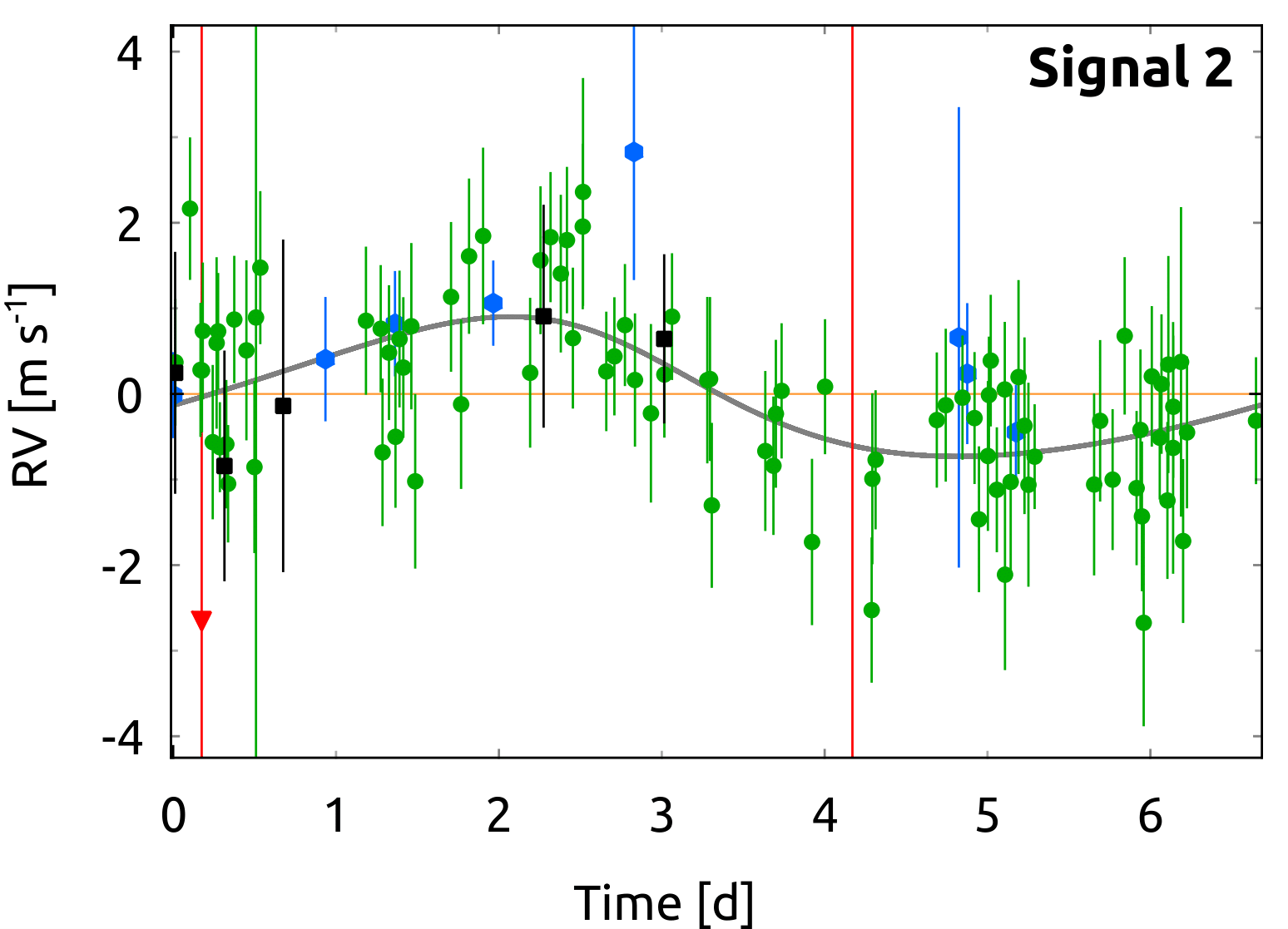}
    \includegraphics[width=0.48\linewidth]{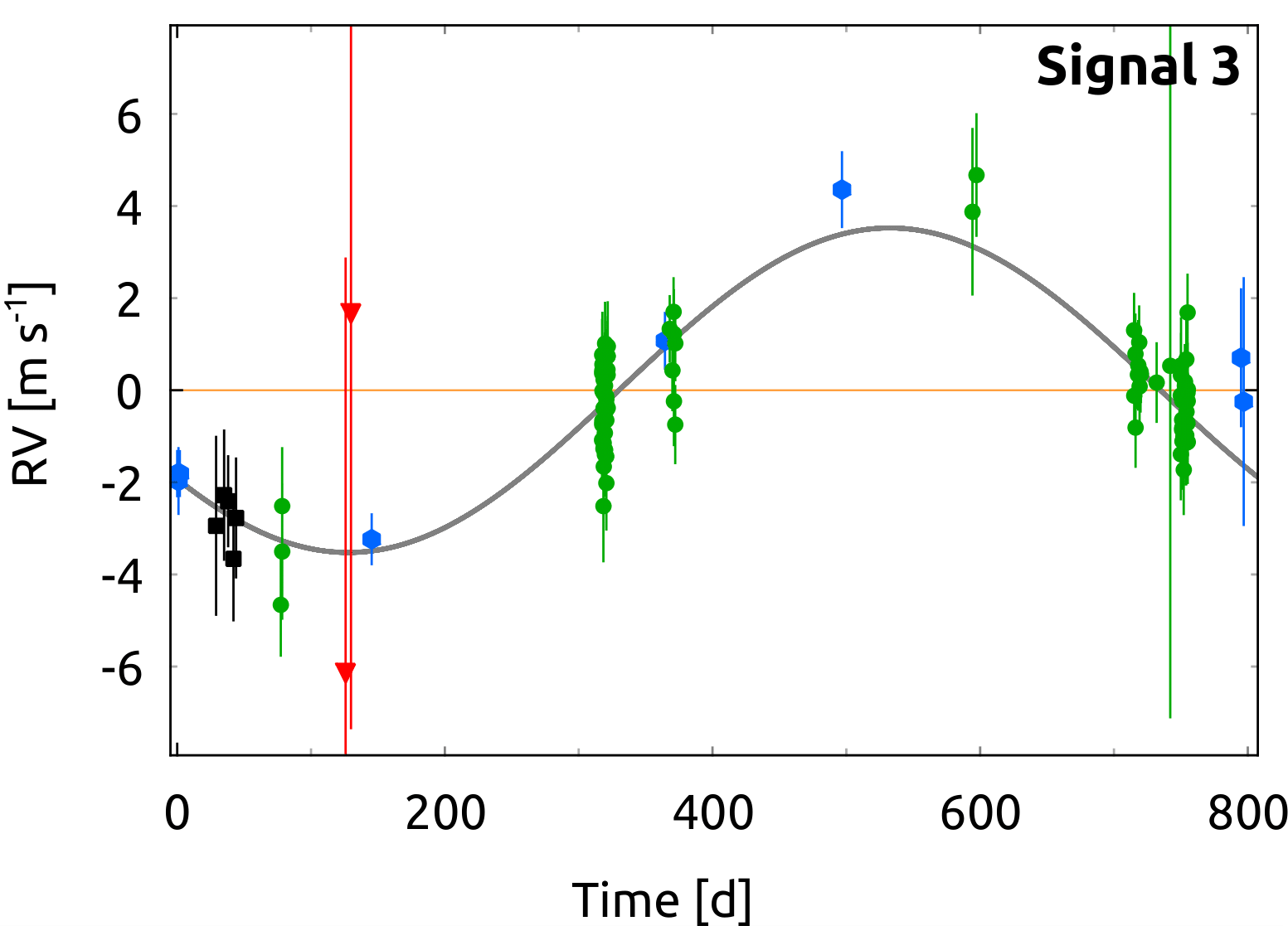}\includegraphics[width=0.48\linewidth]{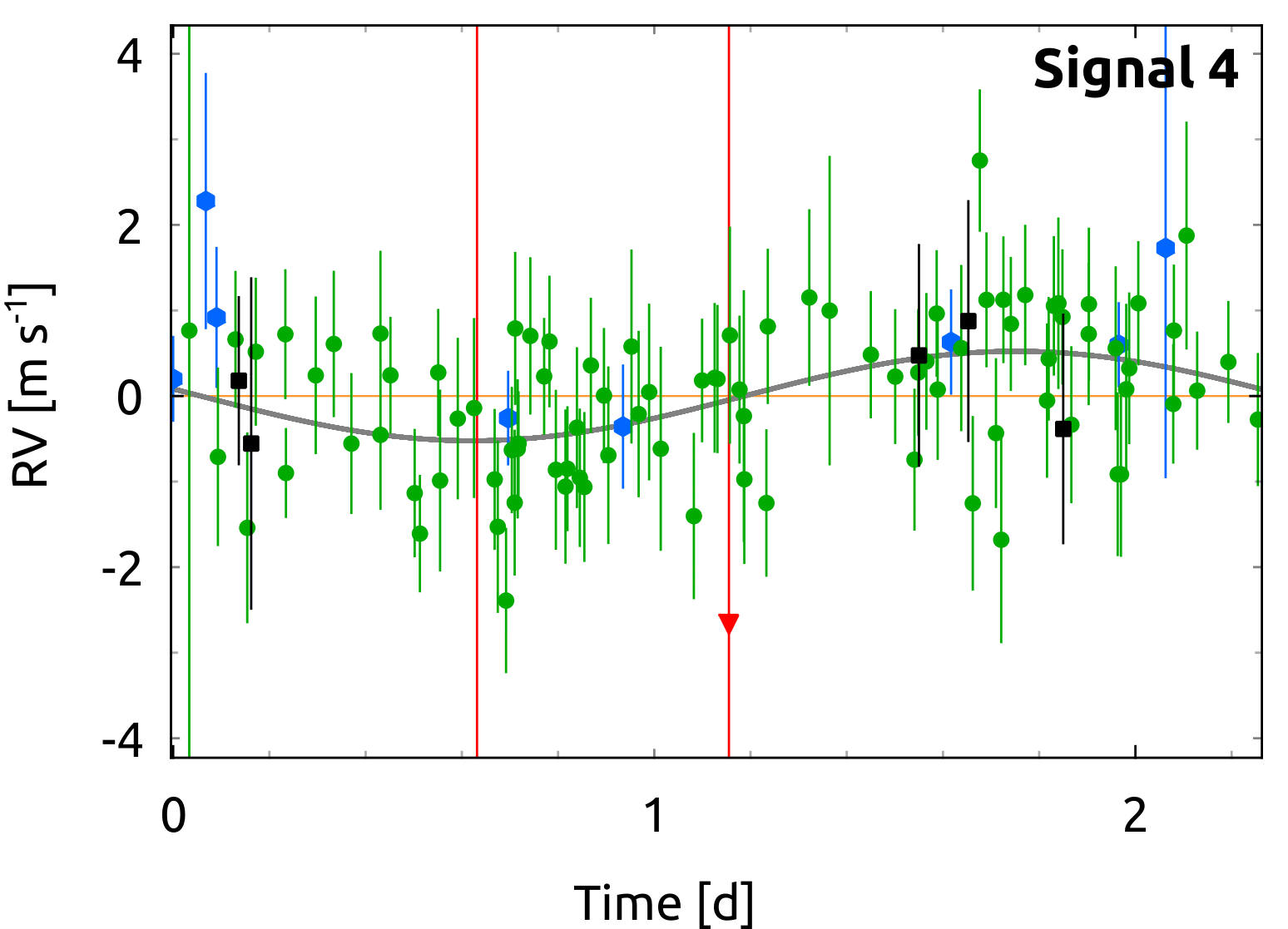}
    \caption{\textbf{Top:} Radial velocities of DMPP-3A. The RVs and residuals are shown along with the model fit for the 4 Kelperian solution described in Table~\ref{tab:objects_params}. \textbf{Middle left:} Phase folded plot of Signal 1, showing the DMPP-3AB binary orbit RVs (with an inset panel showing the S21 data points around periastron passage). \textbf{Middle right:} The phase fold created from the Keplerian solution for a MLE fit at a period of 6.67 d. \textbf{Bottom left:} The RV phase fold of Signal 3 with a folded period of 809~d from MCMC analysis. \textbf{Bottom right:} Phase folded residuals, showing Signal 4. The best-fitting period is 2.26~d. By observing BIC changes it was determined that there was no statistical significance of fitting eccentricity as a free parameter, which is forced here to be e = 0.  \textbf{All plots:} RV folds were created for each Signal with all other identified Signals subtracted off, to show the individual solutions. Archival CHEPS measurements are marked with dark blue hexagons, CORALIE data points are red triangles, the main body of DMPP observations are green circles, and the S21 observations are black squares.}
    \label{fig:RVplots}
\end{figure*}

The DMPP-3AB binary RV modulation (hereafter Signal~1) is shown in Fig.~\ref{fig:RVplots}. The S21 observations with {\sc HARPS} confirm the previous fit and provide tighter constraints on the solution. The new binary parameters and derived system parameters (using our new estimates of the stellar properties from Table \ref{tab:dmpp-3aprops}) are listed in the Signal 1 column of Table~\ref{tab:objects_params}. The significance of the binary orbit, quantified by the Bayesian information criterion (BIC), is improved by three orders of magnitude. The uncertainties in all the parameters are reduced, in some case by orders of magnitude, compared to those in \citetalias{Barnes2020}. The 3$\sigma$ lower limit on the mass of DMPP-3B is now 80.9\,$M_{\textrm{jup}}$ due to increased semi-amplitude and our recalculated mass of DMPP-3A. This suggests DMPP-3B is indeed massive enough to sustain hydrogen burning, and is an object located at the very bottom of the main sequence.

\begin{figure}
\includegraphics[width=0.88\columnwidth]{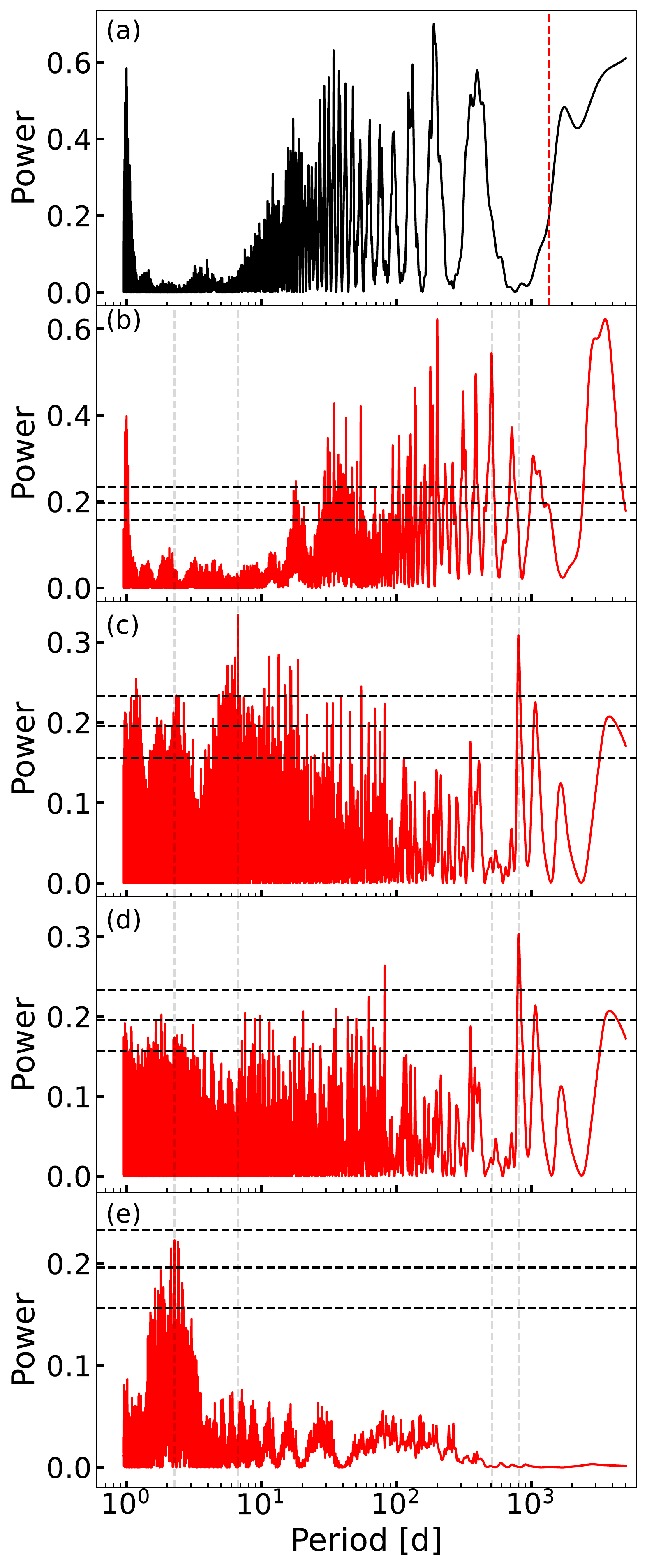}
    \caption{\textbf{(a):} Spectral window function computed by taking the discrete Fourier transform, with a vertical line plotted at a period of 1356~d (see Section~\ref{subsec:800d}). \textbf{(b):} The GLS periodogram of the radial velocities, identifying a $\sim507$~d period corresponding to Signal 1. \textbf{(c):} The 507~d Keplerian has been removed, clearly showing significant peaks at 6.67~d (for the planet) and $\sim800$~d. The other significant peaks in the 10--100 d range are likely a result of aliasing and are in a very busy region of the window function, so are neglected for signal fitting.
    \textbf{(d):} Upon removal of the 6.67~d planet RVs by fitting Signal 2, the periodicity at $\sim800$~d remains. \textbf{(e):} After fitting for Signal 3, a periodicity at $\sim2.26$~d is found in the residuals, at $\sim$~0.2 per cent FAP. In all periodogram plots the dashed horizontal lines correspond to 0.1, 1 and 10 per cent FAP levels, where FAP gets smaller with increasing power. The vertical lines overlaid correspond to the periods of signals listed in Table~\ref{tab:objects_params}, identifying periodogram peaks.}
    \label{fig:RV_GLS}
\end{figure}

Fig.~\ref{fig:RV_GLS}\,c shows the periodogram of the RV data after subtracting our best-fitting binary orbit. The 6.67~d super-Earth planet \hbox{DMPP-3A\,b} (hereafter Signal~2), is resoundingly detected. The phase-folded RV curve of DMPP-3A\,b is shown in the middle right panel of Fig.~\ref{fig:RVplots}. For the parameters of DMPP-3A\,b (Table~\ref{tab:objects_params} Column 3), most values remain close to those previously reported. Period and semi-major axis are both consistent with previous results. $P$ uncertainty remains the same, whilst uncertainty on $a$ is reduced by a factor of 4. The value for $K$ has reduced from 0.97 to 0.82~m\,s$^{-1}$, with uncertainties an order of magnitude smaller. $M_{\textrm{0}}$ and $\omega_{\textrm{0}}$ have changed by $\sim$ 30\,$^{\circ}$, with much smaller uncertainties now as they previously covered a wide range. The eccentricity is changed slightly, at $e=0.17$ compared with $e=0.14$ as found by \citetalias{Barnes2020}. We started with the value of 0.14, and used a uniform prior of $0 \leq e \leq 0.8$ during maximum \textit{a posteriori} fitting and MCMC simulations. The difference in the BIC between circular and eccentric solutions is $\Delta$\,BIC $\sim5$, indicating the data are sufficient to constrain $e$. The uncertainties in $e$ are broadly similar to those of \citetalias{Barnes2020}, with eccentricity value ranges consistent between both analyses. $M_{\textrm{p}}\,\sin\,i$ is reduced as a result of a smaller fitted $K$ value, changing from 2.58 to 2.22~${\rm M_\oplus}$, with uncertainties remaining the same order of magnitude at roughly $\pm$0.5~${\rm M_\oplus}$

There is some variation in a few of the statistical parameters. The RV offset ($\gamma$) values are different from those reported in \citetalias{Barnes2020}. The {\sc HARPS} data were re-reduced with the {\sc terra} software as a single dataset, leading to different systematic offset values. The sampling of periastron has slightly changed the offset parameters too, but only has a small effect. The stellar jitter parameters are reduced slightly, except for $\sigma_{\textrm{CORALIE}}$ which is now small instead of non-zero as previously reported \citepalias{Barnes2020}.

\citetalias{Barnes2020} also identified a periodogram power peak at $\sim$800~d in the RV residuals after fitting for the signals corresponding to DMPP-3B and DMPP-3A\,b, at 6~per~cent false alarm probability (FAP). They tentatively attributed this to intrinsic stellar activity and active regions on DMPP-3A. If dynamical and co-planar, this RV signal would indicate a body on an orbit crossing that of DMPP-3B. A longer temporal baseline now allows us to investigate the cause of this signal (henceforth Signal 3). The addition of the S21 RV data affirms both Signals 2 and 3 in the residual GLS peaks at 6.67~d and $\sim$800~d (Fig.~\ref{fig:RV_GLS}\,c). The phase fold for Signal 3 is shown in Fig.~\ref{fig:RVplots}. This strengthens the argument that Signal 3, with maximum {\it a posteriori} fitted period of $809$~d, is genuinely present and persists throughout the entire timespan of the data. The semi-amplitude of the variation is 3.5~m\,s$^{-1}$. If attributed to a Keplerian orbit, the best-fitting period corresponds to a semi-major axis of approximately 1.64~au.

\subsection{Long period RV variation}\label{subsec:800d}
 
Despite producing a relatively convincing phase fold (Fig.~\ref{fig:RVplots}, bottom left panel), it is difficult to attribute Signal 3 to a body orbiting the host star. If Signals 1 and 3 are co-planar, the orbits would cross due to the large eccentricity of the DMPP-3AB binary orbit (as shown in Fig.~\ref{fig:orbits}). We have performed initial orbital integrations to confirm that in the coplanar case, a putative body following the $\sim$800~d orbit is either ejected or cannot be described with continuous variables. As a further test, we report dynamical simulations relaxing the assumption that the orbits are co-planar in Section~\ref{sec:3Dsims}, performing a more thorough analysis of potential orbital configurations. 
As a consequence of the simulation results, we suspect that this signal is due to activity, and this is investigated in Section~\ref{sec:activity}. A further point to note is that it is unlikely that this confusing signal is a direct artefact of aliasing, as the window function (Fig.~\ref{fig:RV_GLS}\,a) has a substantial local minimum at the period of the signal. 

Alias periods occur due to convolution between the uneven temporal sampling of observations and other physical periods present in the data. The window function is calculated by taking the discrete Fourier transform (DFT) of the time samples. To accomplish this, we first create a frequency array which depends on the smallest and largest frequencies to be used, as well as the step size between samples. For each frequency, an array of observation phases is calculated. Sums of the sine and cosines for every phase are then combined to determine window function power at that frequency. Peaks of this function denote frequencies that cause the largest interference on physical signals.
Alias frequencies (or periods upon inversion) are calculated simply using $f_{\textrm{alias}} = f \pm f_{\textrm{window}}$\,, where $f$ is the peak of the RV periodogram, and $f_{\textrm{window}}$ the peaks in the window function \citep{2010..Dawson}. Taking Signal 1 as the dominant period that could be convolved with sampling, and Signal 3 as the potential alias period, we can rearrange the previous equation to determine where a window function peak would be required for Signal 3 to indeed be an alias of the binary orbit: 
\begin{equation}
    \frac{1}{809.38} = \frac{1}{506.89} \pm \frac{1}{P_{\textrm{window}}}~~[1/\textrm{d}].
\end{equation}

This gives a required window function peak at 1356\,d. There is no peak at this period in the window function (see Fig.~\ref{fig:RV_GLS}\,a): aliasing seems unlikely to cause Signal 3. 

\begin{figure}
    \centering
    \includegraphics[width=0.90\columnwidth]{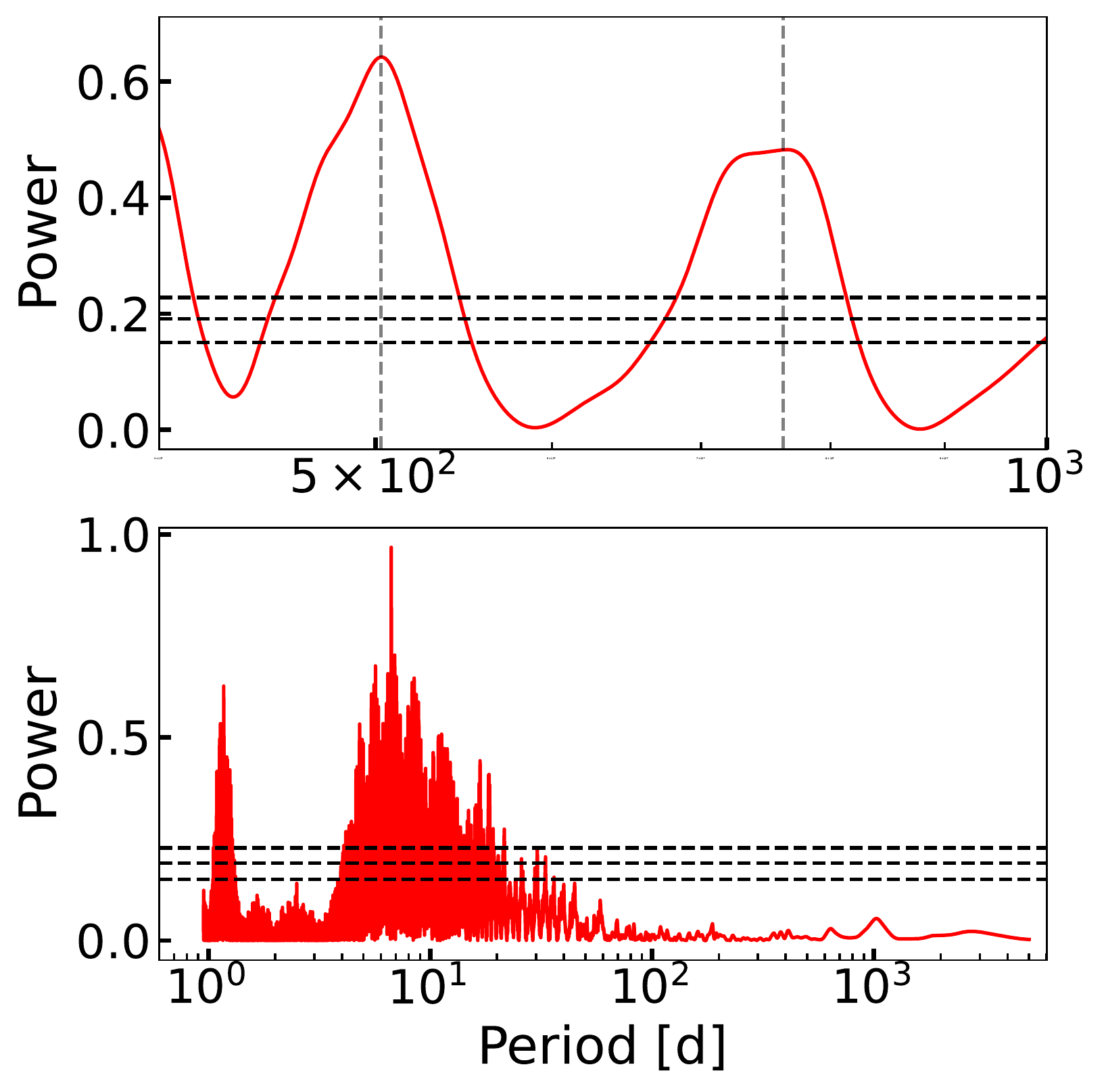}
    \caption{GLS periodograms for investigation into sampling. \textbf{Top:} The periodogram for simulated RVs of DMPP-3B and DMPP-3A\,b. This plot shows a similar but much broader peak in the 700--800~d region than Fig.~\ref{fig:RV_GLS}\,b, with a maximum power at $\sim$ 760~d (denoted by a vertical dashed line, as is the binary period at $\sim$ 507~d). \textbf{Bottom:} After fitting for the binary orbit, the simulated 6.67~d period is by far the strongest peak in the periodogram. There are no long period peaks, and if Signal 3 was caused by sampling, we would expect to see a feature at 800~d (and this plot would mimic an idealised version of Fig.~\ref{fig:RV_GLS}\,c).}
    \label{fig:sim_GLSs}
\end{figure}

While there is no window function peak at 1356\,d, there is power there.
To investigate whether Signal 3 could be an artefact of the temporal sampling of the DMPP-3B orbit, we simulated RVs for Signal 1 {\em only} at the observation epochs of our data. The GLS of the simulated data shows a broad period identified at $\sim$760~d that could be an alias (Fig.~\ref{fig:sim_GLSs} Top panel). By comparison, there is a narrower peak in the observed RV data (Fig.~\ref{fig:RV_GLS}\,b) at a similar period. We next simulated the reflex RV signals of both the binary and the planet DMPP-3A\,b. After fitting and subtracting the binary orbit, the residual GLS periodogram only shows a significant peak at 6.67~d (Fig.~\ref{fig:sim_GLSs} Bottom panel), rather than at 6.67 and 800~d which we would see if Signal 3 was due to sampling. 

We thus confidently rule out the possibility that Signal 3 is created by the temporal sampling of our observations. It is a genuine signal.

As another test of RV signal nature, we investigated the stacked Bayesian GLS periodogram (s-BGLS). Introduced by \citet{2017A&A..Mortier...sBGLS}, this involves computing Bayesian GLS periodogram \citep{2015A&A...573A.101Mortier..BGLS} for an initial number of observations, and then incrementally including successive observations to the GLS computation. With this method we can check the coherence of a signal. A planet signal should be coherent and continually increasing in significance, whereas activity signals (which are often quasi-periodic) should be unstable and incoherent.

\begin{figure}\label{fig:s-BGLS}
	\includegraphics[width=0.99\columnwidth]{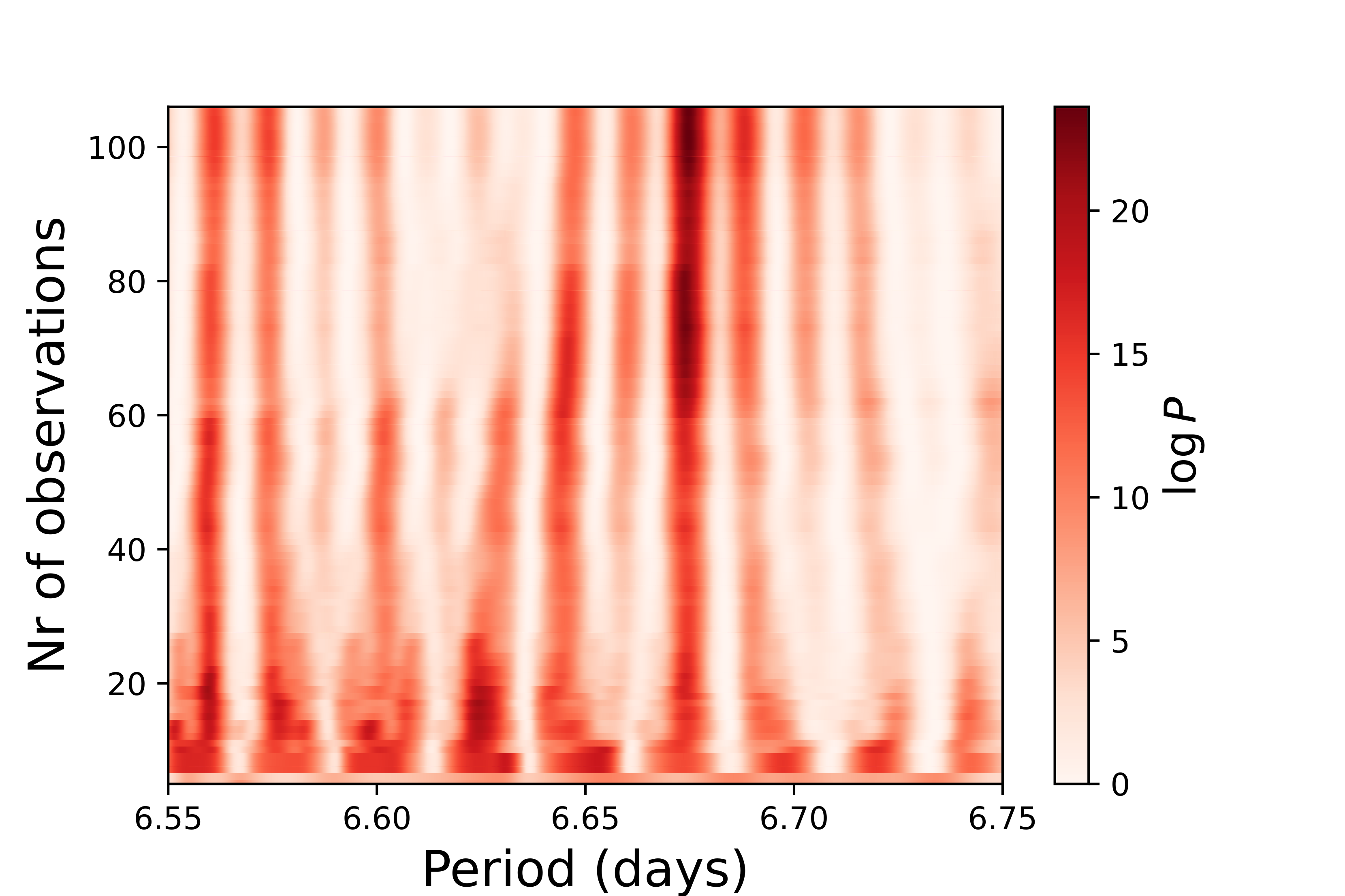}
	\includegraphics[width=0.99\columnwidth]{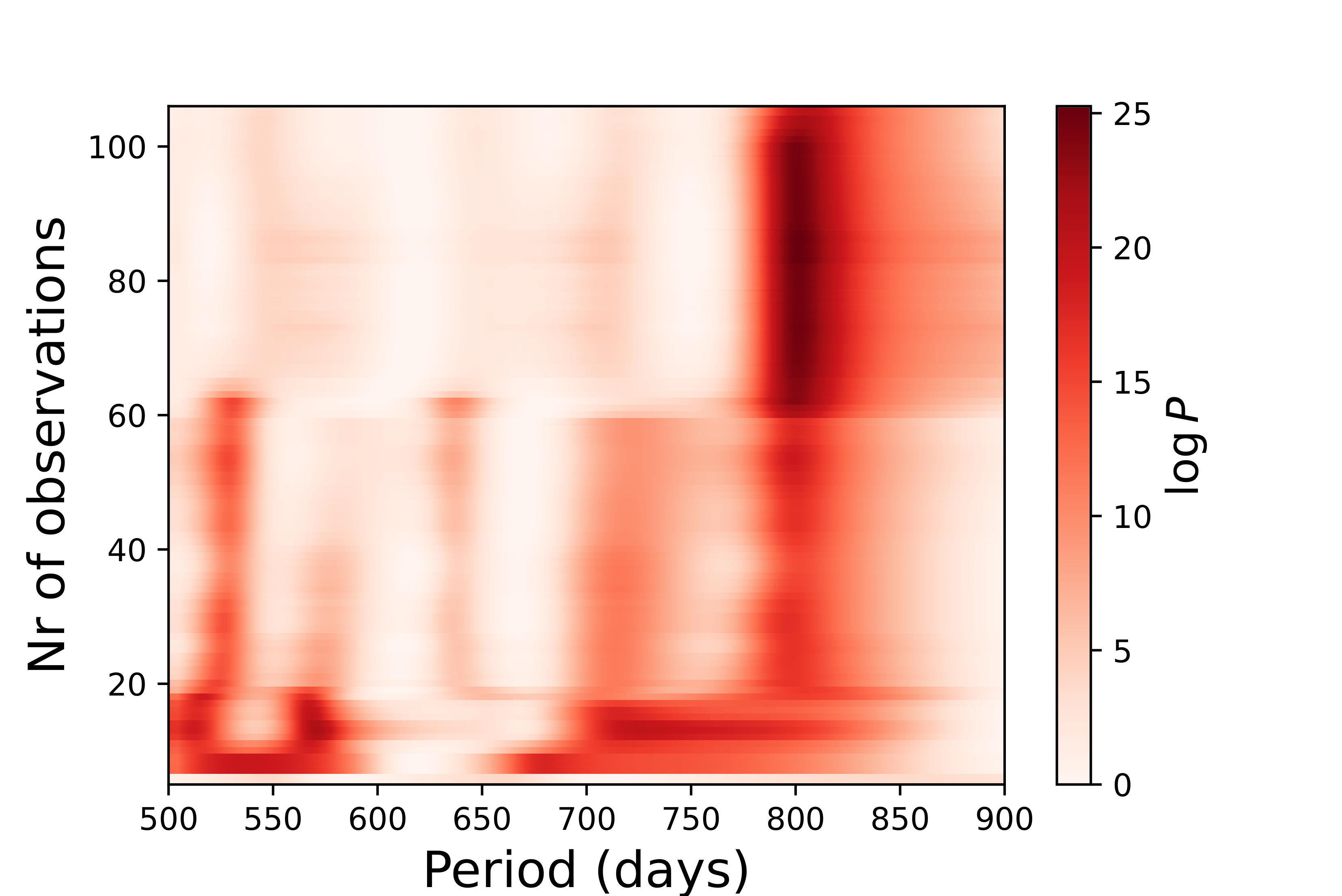}
	\caption{s-BGLS plots, where the Bayesian GLS periodogram is incrementally calculated, to check the coherence of a signal. At the bottom of the plots there is limited data so the periods are not sharply defined. At the beginning of new observing runs there are step changes in the window function, e.g. around observation 60.
    \textbf{Top:} A section of the RV residual s-BGLS after Signal 1 has been removed, focused on the 6.67~d period (with closely spaced, lower power aliases). The coherence of the signal monotonically increases as observations are added.  \textbf{Bottom:} The s-BGLS section around 800 days. The identified peak is not very coherent, and reaches a maximum that decreases slightly with the addition of the final 5 (S21) observations. The 800\,d period is not fully sampled until around observation 20.
    } 
    \label{fig:sBGLS}
\end{figure}

We see that Signal 2 (Fig.~\ref{fig:sBGLS} top panel) is coherent and grows monotonically in strength as one would expect for a genuine planet. The s-BGLS for Signal 3 (Fig.~\ref{fig:sBGLS} bottom panel) also seems to show coherence from observation 20 onwards, after the period of the signal has been sampled. As the observing runs are much shorter than this period, the peak is broad. However, with the addition of the S21 data the significance decreases slightly, and the peak power shifts in period by $\sim2.5$~d. As these observations are separated from the initial DMPP dataset by 3 years, any signals caused by activity could evolve in the intervening time, whereas planetary signals are strictly periodic. The change in significance and period after $\sim100$ observations suggests that Signal 3 is caused by activity. We discuss this in later sections.

\subsection{An additional interior S-type planet?}\label{sec:2dplanet}
We fitted a circular Keplerian to the RV residuals after the subtraction of Signals 1 and 2, thus removing the 
$\sim$ 800~d Signal 3.  The resulting residuals were searched for further periodic modulations. The GLS periodogram shows a $2.26$~d signal, hereafter Signal 4 (see Fig.~\ref{fig:RV_GLS}\,e). This is a tentative detection at $\sim0.2$~per cent FAP. The inclusion of this signal improves the BIC by just over 2, the threshold for positive evidence \citep{RafteryBIC}. This BIC improvement corresponds to a $\sim2$-sigma detection, hence our current tentative stance on the planetary nature of this signal (see Table~2 in \citet{Standing2022}, where $\Delta$~BIC = 2 is equivalent to a Bayes Factor = 3).
The log likelihood improves by $\sim10$, meaning this statistic shows significant evidence that this signal is present in the data (with the threshold for strong evidence being $\Delta~\lnL~>7$, \citealt{Kass+Raftery}). Fig.~\ref{fig:RVplots} (bottom right panel) shows the residual RVs folded on the best fitting solution. The posterior orbital period and other parameters for Signal 4 are given in Table~\ref{tab:objects_params}. 

After fitting and subtracting Signal 4, no further signals remain in the GLS periodogram at FAP < 10 per cent. Signal 4  corresponds to a putative planet orbiting at 0.033~au from DMPP-3A, interior to the planet DMPP-3A b. 

The stability of this two planet configuration has been validated with the \textsc{exo-striker} package, using the symplectic massive body algorithm (SyMBA) functionality \citep{SyMBA1998}. This orbital integrator  allows for the time step to be recursively reduced upon close interactions between bodies in the system to fully simulate the interaction - whilst retaining the speed of traditional mixed variable symplectic (MVS) algorithms, the previous gold standard of orbital simulations \citep{MVS...1991AJ....102.1528W}. 

There is no indication that Signal 4 is due to aliasing: a 1-day alias of DMPP-3A\,b would lie at 1.18~d. The stellar activity and line profile indicators show no periodicities on $\sim2$~day timescales. This is as expected because the timescale is too short to be induced by stellar rotation in an old star, but too long to be attributed to stellar p-mode oscillations \citep{2019.Collier.Cameron,2021MNRAS...Costes}. 

\subsection{Combined RV and Astrometry}\label{sec:gaia}

The reflex motion of a star caused by companions (either planetary or stellar) can be assessed with other techniques aside from radial velocities. Astrometry involves measuring the positions of objects very accurately, and with this we can map out the 3D spatial and velocity distribution of stars. If accompanied by another body, the positions of a host star will vary as a result of motion about the common barycentre, indicating a perturbing presence. 

The cutting-edge of stellar astrometry measurements comes from the third \textit{Gaia} release (DR3; \citealt{GaiaDR3paper}). \textit{Gaia} has been in operation for $\sim$34 months, and therefore has amassed sufficient observations to make the search for binary stars with \textit{Gaia} possible for the first time. 

Stars in the \textit{Gaia} catalogue are considered for multi-star analysis when their motion does not fit the single star astrometric models. Data processing identified these stars through use of the renormalised unit weight error (RUWE) statistic. This is effectively a goodness-of-fit metric, quantifying how well a star can be described as a single body considering the observed movements through space \citep{Almenara2022}. For RUWE $\gtrsim 1.4$, the motion is not consistent with a single star, with the excess astrometric noise potentially indicating an unseen companion \citep{Lindegren2021,Almenara2022}. DMPP-3 has RUWE = 6.85, so it is likely that Gaia has detected significant deviation from a single star solution, and should be able to quantify the motion of the binary system via astrometric measurements. 

The orbital solutions for non-single stars in DR3 are published online \citep{GaiaDR3paper}, and using object identifiers, the parameters can be extracted for particular systems. For stars such as DMPP-3, which have both astrometric and single-lined spectroscopic solutions, the orbital parameters are expressed in the Thiele-Innes elements ($A,B,F,G,C,H$). We have used the conversions detailed in the appendices of \citet{Halbwachs2022} to convert these to the traditional Campbell orbital elements ($a,\omega,\Omega, i$), along with associated errors.

Using this conversion, we can determine the inclination of the binary system DMPP-3AB with respect to the observers line of sight. This allows us to eliminate the $\sin{i}$ term in the projected mass retrieved from RV analysis. By assuming that the binary and planet(s) orbit in the same plane, we can also form estimates for `coplanar' masses of any bodies orbiting the central star. This adds credibility to the super-Earth mass of DMPP-3A\,b, and is further evidence for a rocky composition. These masses are listed in Table~\ref{tab:gaia}.

Astrometry also measures the semi-major axis of the primary star's orbit about the system barycentre, $a_{\textrm{A}}$ \citep{Ranalli2018, Halbwachs2022}. Calculated from Thiele-Innes elements $C$ and $H$, we find this to be $0.093 \pm 0.005$~au (Table~\ref{tab:gaia}). This is in excellent agreement with that derived from RV analysis (0.094 $\pm$ 0.001~au). We observe almost negligible deviation in the orbital sizes, at a level below the error bounds in the parameters.

\begin{table}
	\centering
	\caption{Gaia derived parameters, and derived masses constrained by inclination and the assumption that all bodies lie in the same orbital plane.}
	\label{tab:gaia}
	\begin{tabular}{lc}
		\hline
		Parameter & Value \\
		\hline
        $a_{\textrm{A}}$~(au) & $0.093 \pm 0.005$ \\
		$i$~$(\degr)$ & $63.89 \pm 0.78$ \\
		$M_{\textrm{B}}$~(M$_{\textrm{jup}}$)  & $91.90 \pm 0.85$ \\
		$M_{\textrm{A\,b}}$~(M$_{\earth}$) & $2.47 \pm 0.56$ \\
		$M_{\textrm{Sig.4}}$~(M$_{\earth}$) & $1.186 \pm 0.289$ \\
		\hline
	\end{tabular}
\end{table}

\section{Dynamical Simulations of Mutually inclined Orbits} \label{sec:3Dsims}

\citetalias{Barnes2020} discussed only the case that the orbits corresponding to Signals 1 and 3 are coplanar. Since DMPP-3 is an exotic system that challenges current understanding, we cannot dismiss the possibility of mutually inclined orbits. We explore the stability of these configurations with orbital integrations performed with the \textsc{rebound} N-body code \citep{2012...Rein...rebound}. In performing dynamical simulations we have used multiple integrators to ensure confidence in our results. The first integrator used was IAS15, a non-symplectic 15th order Gauss-Radau integrator with adaptive time stepping (see \citealt{reboundias15} for further information). The second was WHFast, a symplectic Wisdom--Holman integrator \citep{MVS...1991AJ....102.1528W,reboundwhfast}. A high order kernel was used in WHFast to improve the accuracy of the integrations \citep{reboundhighorder}.

To investigate the stability of the companion star and a hypothetical body orbiting as described by Signal 3, we neglected the circumprimary planet(s): their small mass and close-in orbits mean they have a negligible effect on the large scale system dynamics. 

\citet{SyMBA1998} describe the ideal time step needed for an integrator as $\sim 10^{-3}$ of the smallest orbital period. Accordingly we set the time step for WHFast to 0.5~days. We assumed the orbit of DMPP-3B is edge-on ($i= 90^{\circ}$), and varied the inclination of Signal 3 between $90^{\circ}$ and $270^{\circ}$, to investigate the full range of prograde and retrograde relative orientations. Despite having an estimate of the inclination from \textit{Gaia} astrometry (see Section~\ref{sec:gaia}), we retain the edge-on assumption here. If DMPP-3B inclination is included, the mass would be higher by a factor of $\frac{1}{\sin{i}}$ and the gravitational interactions we are simulating would be stronger. To investigate potential stability, we have therefore chosen to use the very minimum possible mass for DMPP-3B, to rule out even the most feasible scenario for stable orbits. Relaxing this assumption therefore would only strengthen the conclusions we will draw from the simulations. 
We performed simulations for each mutual inclination for a set of 100 configurations: a grid of $10 \times 10$ evenly spaced starting points for both DMPP-3B and the potential Signal 3 object. Signal 3's orbit was assumed to be circular, as there is no evidence for non-zero eccentricity.

\begin{figure}
\centering
	\includegraphics[width=0.85\columnwidth]{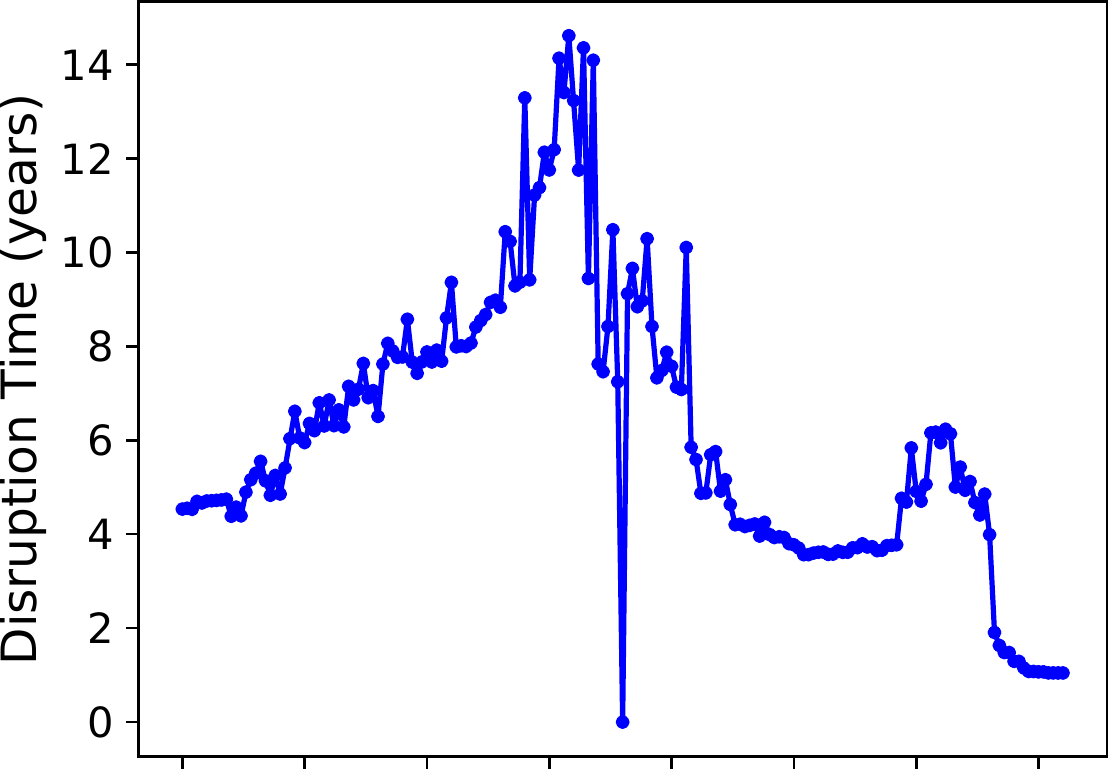}
	\includegraphics[width=0.85\columnwidth]{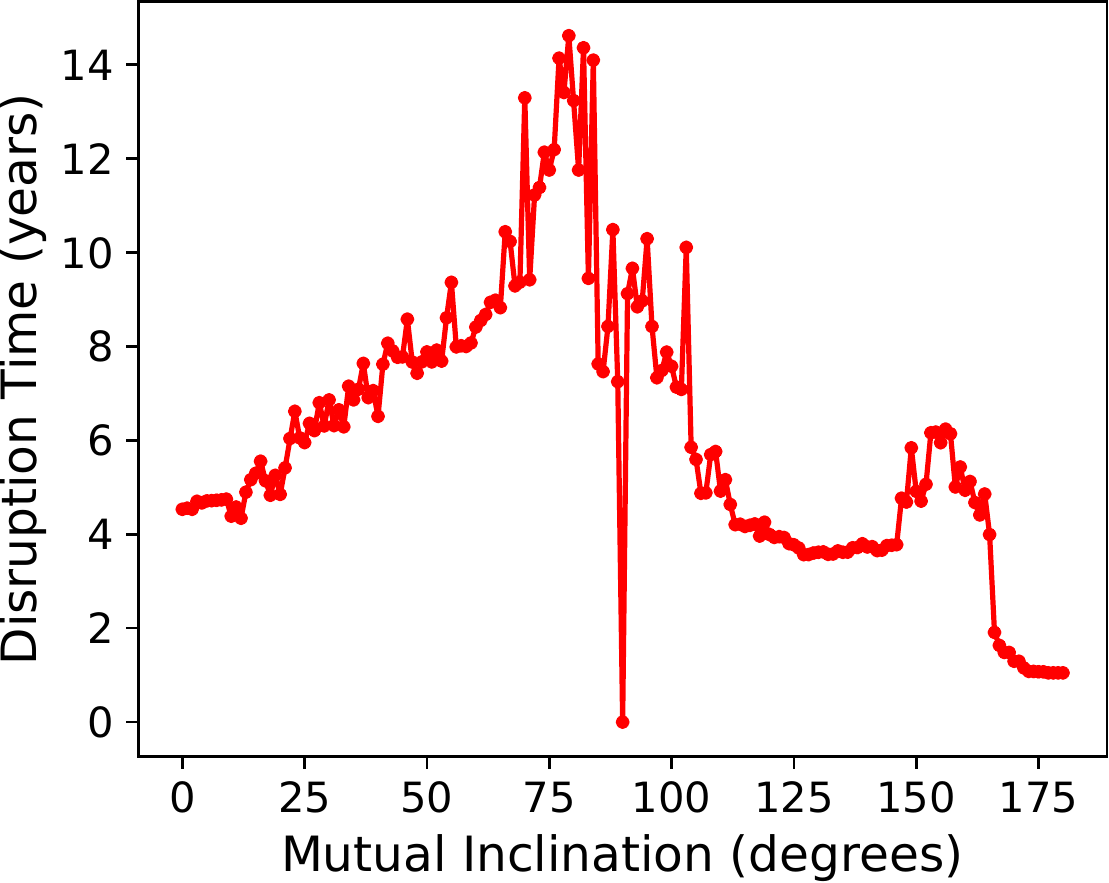}
    \caption{Figure showing median time (for 100 different set-ups) until system disruption, the time where eccentricity passes the value $e = 0.98$, indicating the orbit becomes unbounded. This is calculated for different orbital configurations, showing variation with mutual inclination between the two objects. The inclination for DMPP-3B is set at 90$^{\circ}$, and the inclination of an 800~d period object varied. \textbf{Top:} integrations for this plot are performed with IAS15. \textbf{Bottom:} integrations performed with WHfast. All simulations predict instability at very short timescales. The two integrators give the same result, verifying that different truncation errors in simulations are not causing the observed timescales.}
    \label{fig:inclined_orb_evol}
\end{figure}

We present the results of the investigation into dynamical simulations in Fig.~\ref{fig:inclined_orb_evol}. Simulations were generally performed until the eccentricity of the 800~d signal became $>1$. In most cases this happened very quickly (years), so we did not need to extend total integration time to very large values. The evolution was generally chaotic. A few initial set-ups led to long $e<1$ timescales. Upon investigation, these all asymptotically approached being unbound. We therefore pragmatically dealt with this by slightly reducing the criteria for system disruption. We chose to use $e= 0.98$ as the threshold.

The resulting trend in disruption timescales 
can be easily understood. The stability time rises from the co-planar case (where the orbits are sure to cross) up to a maximum at approximately $75^{\circ}$--$80^{\circ}$ mutual inclination. As the Signal 3 orbit approaches face-on, the putative mass asymptotically approaches infinity and thus the mutual gravitational interactions strengthen. 

Mutual inclinations of $90^{\circ}$--$180^{\circ}$~degrees indicate retrograde orbits. In all cases, the disruption time is shorter for the retrograde case than in the corresponding prograde case because close approaches occur more frequently.

The simulations imply there is no stable situation for Signal 3 to be attributed to an orbiting object. The longest median timescale identified in the simulations is $14$--$15$\,yr. This also suggests that it is unlikely the system could exist in this state for a short period and evolve to some other configuration. The observed 800~d modulation appears roughly sinusoidal over the 13\,yr baseline (Fig.~\ref{fig:RVplots}) showing no sign of the chaos and rapidly-induced eccentricity we see in the simulations. We rule out the possibility that Signal 3 is caused by a planet with $P_{\rm{orb}} \sim 800$~d,
and instead focus on assessing whether this modulation can be attributed to stellar activity.

\section{Stellar Activity}\label{sec:activity}
We use the CCF full-width at half-maximum (FWHM), CCF contrast, and bisector inverse slope (BIS) time-series from the {\sc drs} pipeline, as well as the S-index, H\,$\alpha$, and Na\,D from {\sc HARPS} spectra to thoroughly investigate our data for signatures of stellar activity. The atomic emission line strengths are extracted from the spectra following the procedure detailed in \citet{Barnes2016}, and we direct the reader to that work for further information on how these time-series were produced.  

As we explain in Section~\ref{sec:rvanal}, we suspect Signal 3 is a stellar activity artefact. We also wish to check that the signals we attribute to orbiting bodies are indeed of dynamical origin. 

\subsection{S-index}
The S-index is derived from the \ion{Calcium}{ii} \textrm{H} \& \textrm{K} line core emission, and is calculated as described in \citet{2011..Lovis} and \citet{2021MNRAS...Costes}. The S/N for this indicator is not high due to the very low chromospheric emission from DMPP-3 and the poor throughput at the extreme blue end of the HARPS spectral coverage. The average S/N for \textsc{HARPS} spectral order 7 (containing both H\&K lines) in our data is $<15$ (in comparison to S/N of $\sim100$ at redder orders), and will be reduced further for the calcium lines, due to these being at the edges of the order. The blaze function peaks very strongly at the centre of the order, so the flux will be diminished towards the edges.

\begin{figure}
	\includegraphics[width=0.9\columnwidth]{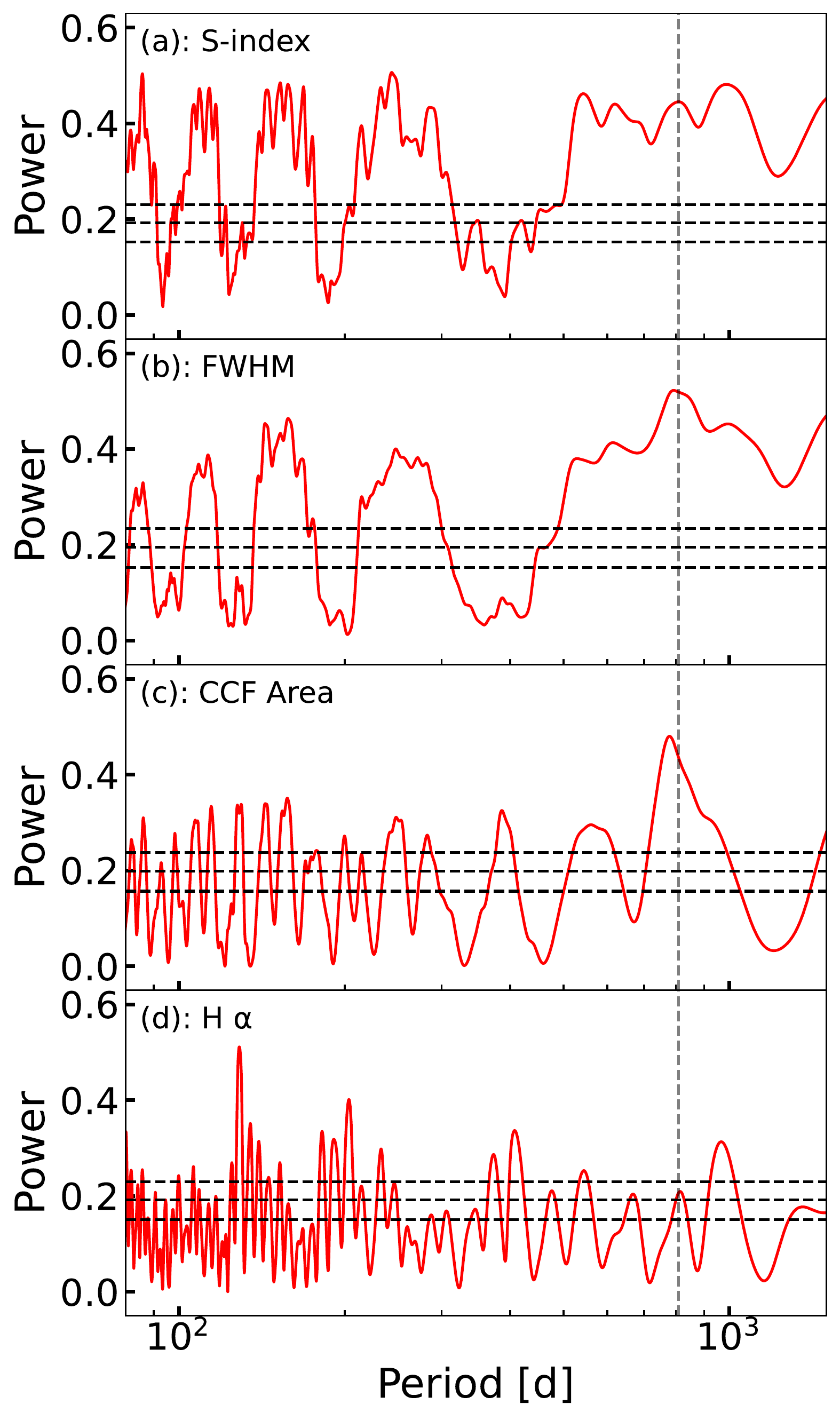}
    \caption{GLS periodograms for activity indicators. The S-index \textbf{(a)}, FWHM \textbf{(b)}, CCF Area \textbf{(c)} and H\,$\alpha$ \textbf{(d)} are shown for the periods longer than $100$~d, with the most significant peaks generally found around $\sim800$~d. The S-index and H\,$\alpha$ both show local peaks that are not the most significant globally, whereas CCF FWHM and Area have stronger connections to the period of Signal 3 - illustrated with a dashed vertical line in all plots. The dashed horizontal lines correspond to 0.1, 1 and 10 per cent FAP levels as before, with these peaks in S-index, FWHM, and Area clearly more significant than the 0.1 per cent level.}
    \label{fig:activity_gls}
\end{figure}

We searched for periodicities in the S-index time-series using the GLS periodogram shown in Fig.~\ref{fig:activity_gls}\,a. There is an extremely broad band of power exceeding 0.1 per cent FAP ranging from around 500-1000\,d which covers the 800\,d period of Signal 3. 
The S-index and RV residuals after subtraction of DMPP-3B's orbit show a very weak negative correlation. The noise present in the S-index time-series may be obscuring the correlation.

\begin{figure}
	\includegraphics[width=0.90\columnwidth]{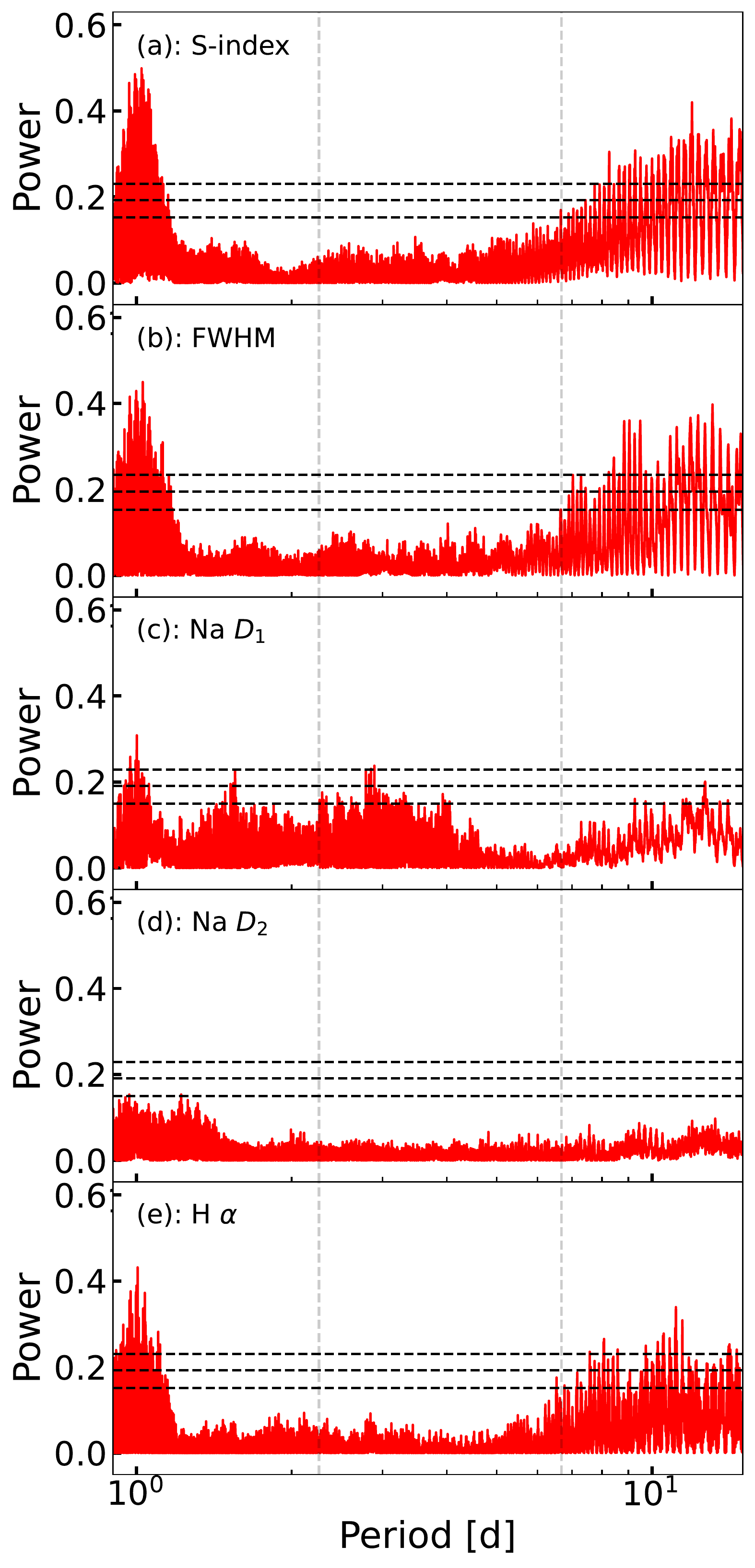}
    \caption{GLS periodograms focused on periods corresponding to S-type planet Signals 2 and 4, with vertical dashed lines indicating periods of 2.26 and 6.67~d. \textbf{(a):} S-index, \textbf{(b):} FWHM, \textbf{(c):} Na~$D_{1}$, \textbf{(d):} Na~$D_{2}$, \textbf{(e):} H~\textalpha. No indicator shows significant periodicities at the relevant period values, with FAP levels denoted by dashed horizontal lines, as before.}
    \label{fig:act_0_10_GLS}
\end{figure}

There is no trace of significant S-index or other activity indicator power at either the 6.67~d period of DMPP-3A\,b, or the 2.26~d period potential planet Signal 4 (Fig.~\ref{fig:act_0_10_GLS}\,a).

\subsection{Full width at half maximum}\label{sec:fwhmanal}
The FWHM of the CCF used to extract the stellar reflex RVs is another identifier we can use to track stellar activity (\citealt{2009AA...Queloz}; Barnes et al. in preparation). 
 
To analyse the FWHM values for any long term periodicities, we must first correct for instrumental effects, see Appendix~\ref{appendix:fwhm}. The period analysis for the corrected {\sc drs} FWHM values gives a GLS periodogram with a peak at $\sim$800~d (Fig.~\ref{fig:activity_gls}\,b), with a best-fitting sinusoid of 792~d period. These values are consistent with Signal~3 (Fig.~\ref{fig:RV_GLS}\,c). We searched for additional periodicities in the residual FWHM time-series after subtracting a $792$~d period sinusoid. None were found. Through our extension of the baseline and a formal treatment of the CHEPS archival measurements, the $\sim800$~d FWHM periodicity identified by \citetalias{Barnes2020} has been refined: the peak has become sharper, with a higher level of significance. 

\begin{figure}
	\includegraphics[width=0.95\columnwidth]{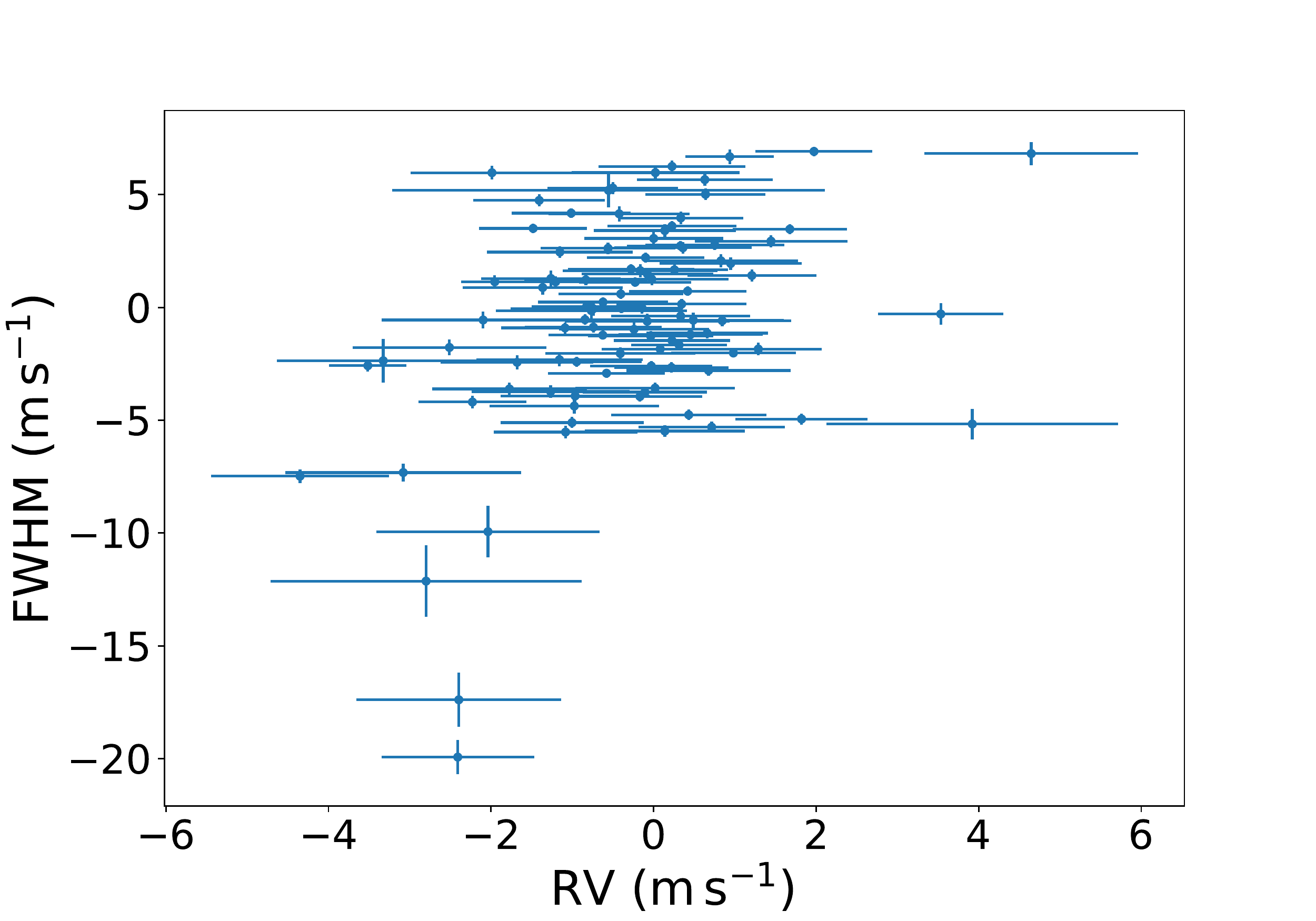}
    \caption{Median subtracted FWHM plotted against RV residuals, after subtraction of Signal 1 \& 2 in Table~\ref{tab:objects_params}. }
    \label{fig:fwhm_rv_corr}
\end{figure}

We inspected the correlation between the FWHM activity indicator and RV residuals, after subtraction of Signals 1 \& 2, shown in Fig.~\ref{fig:fwhm_rv_corr}. The Pearson's $r$ is $0.39$, with $F$-test $p$-statistic of $4.9\times10^{-5}$. This indicates a stronger detection of correlation than reported in \citetalias{Barnes2020}, $r=0.30$ \& $p = 3.7\times10^{-3}$. The connection between stellar activity and the origin of Signal 3 is supported.

We also searched for periodicities in the FWHM to assess the origin of Signals 2 and 4: as with S-index, there is no significant power at the 6.67~d period of DMPP-3A\,b, or for the potential planet Signal 4 at 2.26~d (Fig.~\ref{fig:act_0_10_GLS}\,b).

\subsection{Bisector inverse slope}\label{sec:bis}

\begin{figure}
\centering
	\includegraphics[width=0.98\columnwidth]{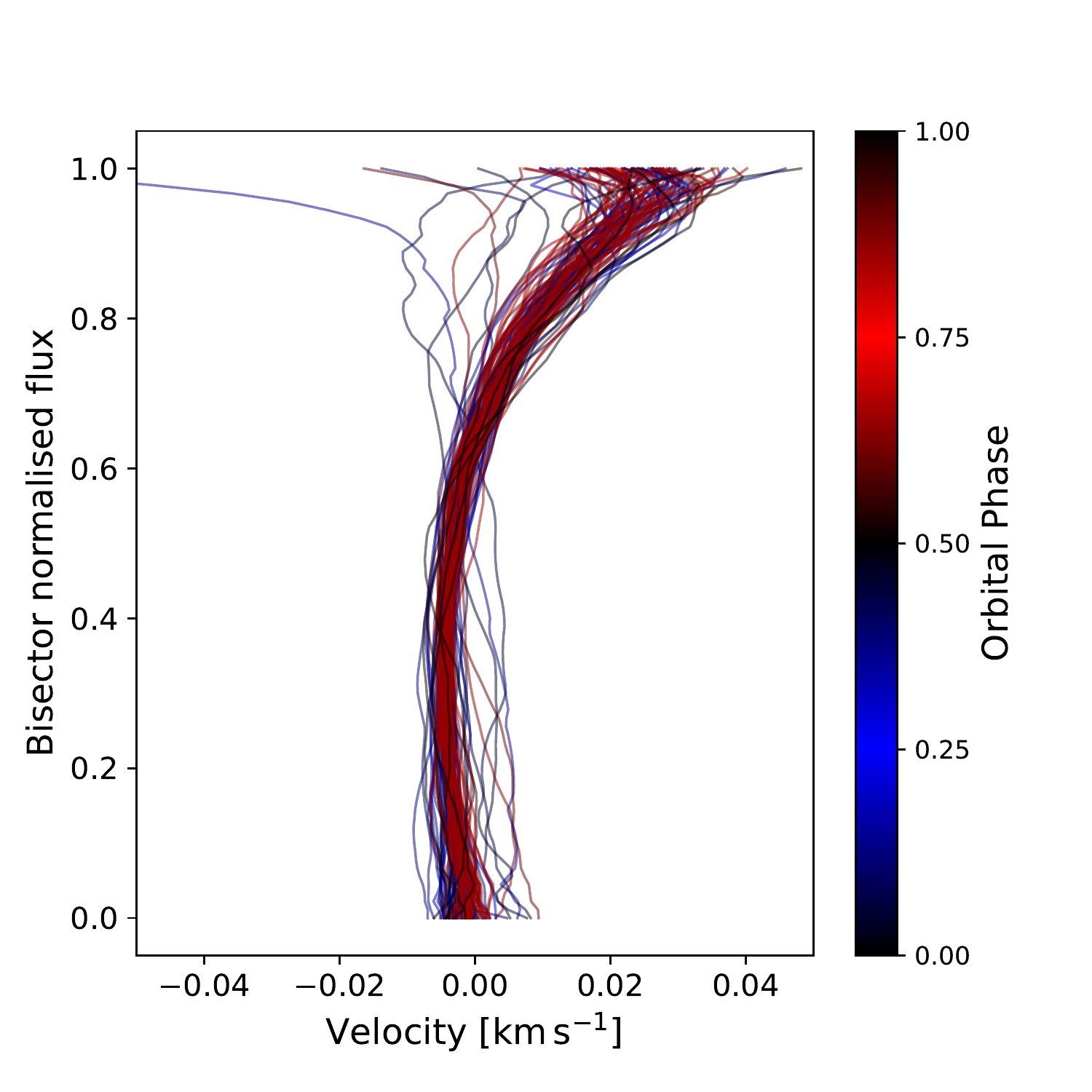}
    \caption{Bisectors of DMPP-3A, colour coded by phase for Signal 3 (Table~\ref{tab:objects_params}). Projected quadrature points of the signal (were this due to a orbiting body) are blue and red, with conjunctions in black. The bisectors are parallel shifted by the barycentric drift corrected RV velocity, output from the {\sc HARPS} {\sc drs}.}
    \label{fig:bisectors}
\end{figure}

We performed periodogram analysis of the BIS time-series which failed to identify any significant periods, with all the peaks below the 10 per cent FAP level. The bisectors appear to be well behaved, with only a couple of observations showing changes in the line profile shapes, which can be seen in Fig.~\ref{fig:bisectors}. 
The most drastically different line shapes correspond to the lowest signal-to-noise (S/N) observations. There is no significant sign of activity in these bisector shapes.
This is consistent with a dynamical origin for Signals 2 and 4, but provides no evidence that Signal 3 is an activity artefact. Importantly, the bisectors are also not modulated on the binary orbital period either, further solidifying the dynamical interpretation of the eccentric binary solution.

\subsection{CCF contrast and area}\label{sec:contrast}
Contrast is defined as the amplitude (or depth in comparison to the continuum) of the inverse Gaussian function fitted to the CCF profile \citep{Gunther2018,2019.Collier.Cameron}. As the intensity of the spectral lines described by the CCF changes, the depth will also change, tracking stellar activity where fluxes are affected by dark spots or bright plages. The measurements of FWHM and contrast are expected to be linked, as changes in the spectral lines should alter both the depth and width of the fitted profile. We do measure a weak negative correlation between FWHM and depth: the data give a Pearson's $r=-0.163$, with a $p$-statistic $= 0.104$

Combining the FWHM with contrast we obtain the CCF area, which is also often used as an activity diagnostic \citep{2019.Collier.Cameron,2021MNRAS...Costes}. The GLS periodogram of this time-series is shown in Fig.~\ref{fig:activity_gls}\,c, and shows a $\sim 800$\,d periodicity with similar significance and a sharper peak than that in the FWHM alone. This is further evidence for a connection between Signal 3 and the stellar line profiles.

\subsection{Periastron events}\label{subsec:periastronevents}

Stellar activity can be enhanced by the tidal and magnetic interactions in close binary stars. With a periastron distance of 0.498~au, DMPP-3 is not a particularly close binary. Since magnetic forces between stars generally decline more rapidly with separation than tidal forces \citep[e.g.][]{2022AJ....164..229B}, the tidal effects will dominate in DMPP-3AB. A number of binary systems show signs of stimulated activity around periastron passage, suggesting a connection with tidal effects induced by the companion star \citep{Moreno2011}. Close binary stars generally show higher levels of chromospheric activity than single stars of the same mass \citep{Eker2008,Qian2012}, and for eccentric orbits this is strongest at periastron \citep{2022MNRAS.515.3716Frasca}. \citet{Moreno2011} study the causes of enhanced activity during periastron events. They find that as tidal shear deforms the stellar surface, energy is dissipated into the stellar layers as heat. The rapid increase in energy deposited around periastron is a promising mechanism to explain increased stellar activity at these orbital phases. 

The manifestation of increased activity is likely to be noticed in emission line changes or enhanced X-ray luminosity during periastron \citep{Moreno2005,2020MNRAS.497..632L}. When the star's surface is perturbed by a companion, patches of the photosphere move at different speeds, and the departure from uniform effective temperatures and gravities induces line profile variability \citep{2016..Harrington}. Inhomogeneity of the external stellar gas layers is evident when observing changes in photospheric lines. The strongest line variation is also seen during periastron, when the influence of the binary companion is strongest.

\begin{figure}
    \centering
    \includegraphics[width=0.95\columnwidth]{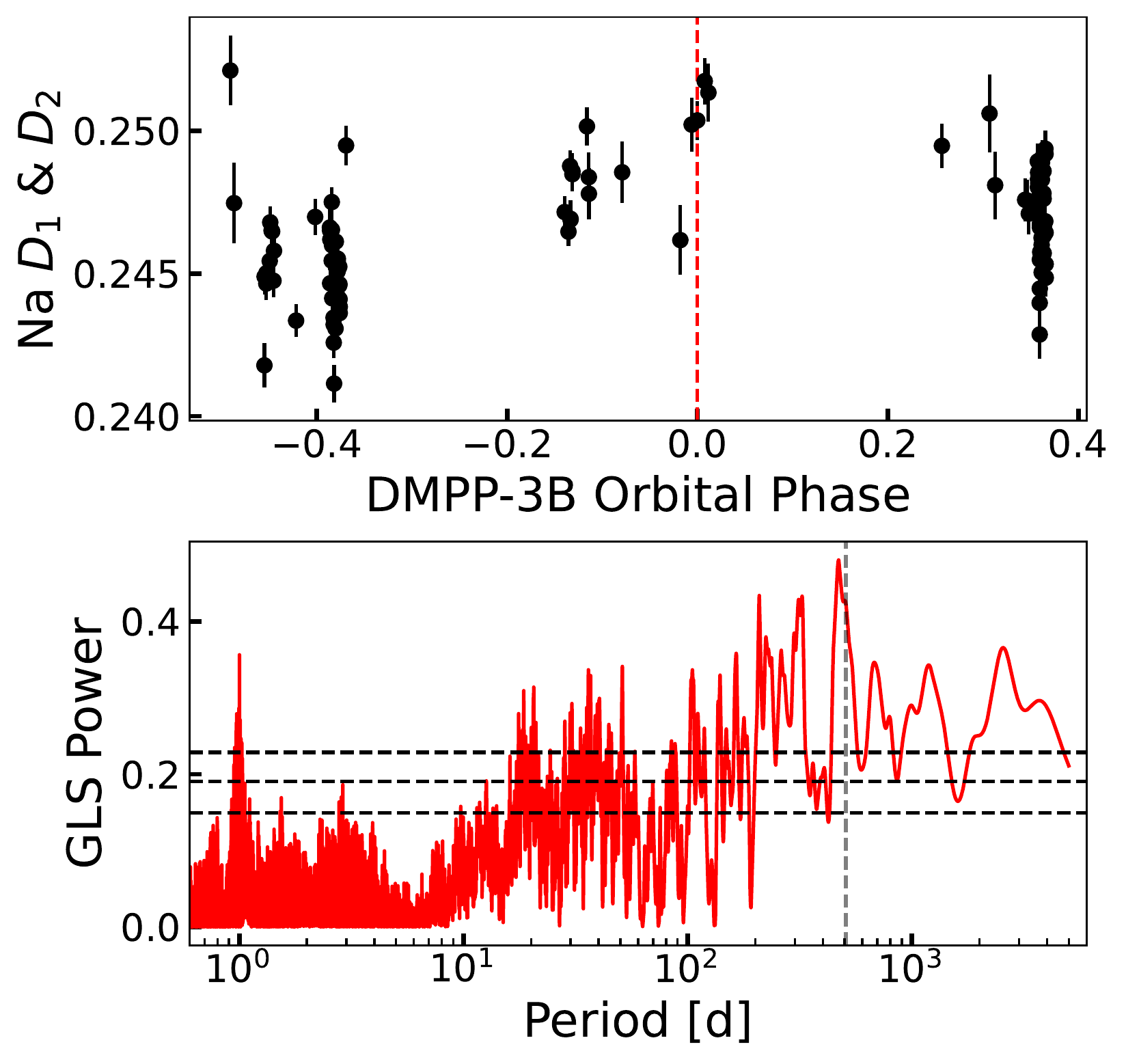}
    \caption{Investigation into enhanced activity around periastron. \textbf{Top:} Combined observations of line strengths for the sodium doublet, phase folded to the binary orbital period (which is at phase $=0$, denoted with the dashed vertical line). We see a slight increase in comparison with the main body of remaining observations. \textbf{Bottom:} The GLS periodogram computed for the sodium time series data, identifying a period near the binary orbit of $\sim507$~d (again marked with a dashed vertical line).}
    \label{fig:NaPeriastron}
\end{figure}

For the configuration of DMPP-3AB we might expect some sign of enhanced activity around periastron. Fig.~\ref{fig:activity_gls} shows that there are no well-defined peaks in most activity indicator periodograms at the binary orbital period. This may be because the modulation is non-sinusoidal. We have searched the time series of activity indicators, to investigate whether we see any enhanced activity. The only indicator that hints at increased activity around periastron is the sodium emission. By combining the two doublet lines, we have inspected the binary orbit phase-folded points and found a slight enhancement around periastron (phase 0, Fig.~\ref{fig:NaPeriastron}, top panel). The periodogram of this time-series, shown in the bottom panel of Fig.~\ref{fig:NaPeriastron}, identifies a global peak relatively close to the $\sim507$~d period of the binary orbit (indicated with a dashed vertical line).
However, it is perhaps over-zealous to attribute the increased sodium emission to an observable activity increase in the DMPP-3 system: \citet{Eker2008} predict that periods longer than a few hundred days will not be important for increased chromospheric activity. 

Future observations will establish whether the elevated Na\,D emission is reproducible. So far, we only have {\sc HARPS} activity indicator information during a single periastron passage. The next periastron epoch is forecast to be in 2023 (around February 09), and observing this would provide a far stronger assessment of periastron effects in the DMPP-3 binary system.

\subsection{Rotation period searches}\label{sec:rotation}
 
Values for projected rotation velocity ($v\sin{i}$) of DMPP-3A vary from source to source, as shown in Table~\ref{tab:rot_vels}.  DMPP-3A is a slow rotator and reliable measurement of $v\sin{i}$ close to or below the instrument resolution is challenging \citep{2007A&A...467..259..Reiners}. {\sc HARPS} has resolution $R$ ($\Delta\lambda/\lambda$) of $\sim 115000$, corresponding to  $\sim$~2.6~km\,s$^{-1}$ so we can rule out $v\sin{i} >> 2.6$~km\,s$^{-1}$ but a precise measurement is difficult and dependent on details of the instrumental line spread function. Nevertheless, by using $P_{\textrm{rot}} = 2\pi R\,\sin{i} /~v\sin{i}$, we can obtain lower limits for $P_{\textrm{rot}}$ assuming $0.5$~km\,s$^{-1}$ $\leq$~$v\sin{i}$~$\leq 3$~km\,s$^{-1}$, c.f. Table~\ref{tab:rot_vels}. We find $15\,\textrm{\,d} \lesssim P_{\textrm{rot}} / \sin{i} \lesssim 90$~d. These is as expected for a star of this spectral type and age \citep{1986PASP...98.1233..Stauffer,2015MNRAS.452.2745..SM,2020AJ....160...90..Angus}.

\begin{table}\label{tab:vsini}
	\centering
	\caption{Projected rotation velocities of the star DMPP-3A (HD42936). These values were obtained from the SIMBAD database, and are listed with references to the publications they originate from. For information on how each rotation velocity was calculated, see individual sources.}
	\label{tab:rot_vels}
	\begin{tabular}{lcccc} 
		\hline
		$v\sin{i}$ (km\,s$^{-1}$) & Reference \\
		\hline
		0.5 $\pm$ 1  & {\citet{2020A&A..Hojjatpanah}} \\
		1.4 $\pm$ 0.1 & \citet{2017MNRAS.468.4151..Ivanyuk}  \\
		1.643 $\pm$ 0.437 & \citet{2018..SPECIES.I}  \\
		1.97 $\pm$ 0.14 & \citet{Barnes2020} \\
		2.7 & \citet{2011AA...531A...8Jenkins}  \\
		3.17 $\pm$ 0.1 & {\sc species} (see Table~\ref{tab:dmpp-3aprops})  \\
		\hline
	\end{tabular}
\end{table}

\begin{figure}
    \centering
    \includegraphics[width=0.90\columnwidth]{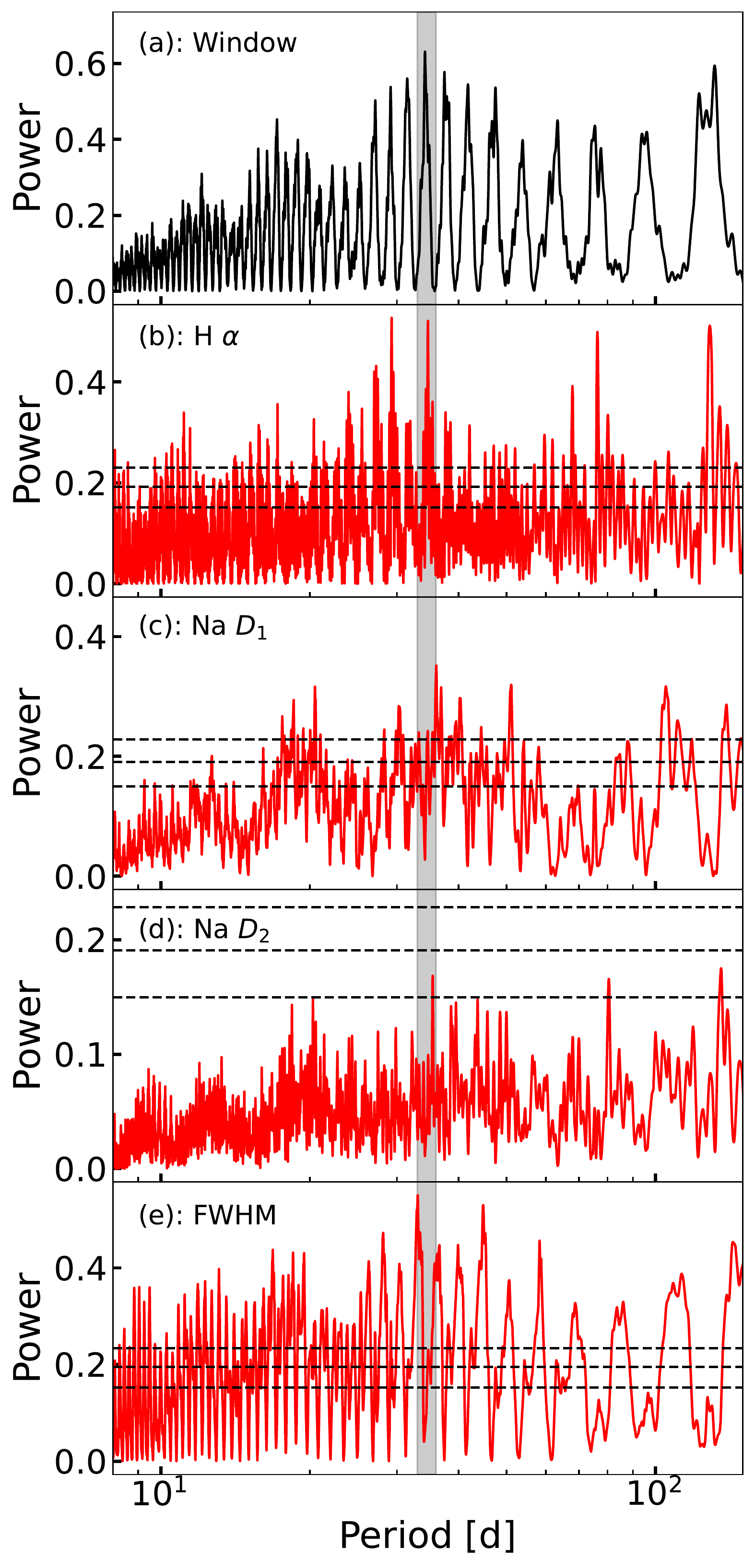}
    \caption{The spectral window function \textbf{(a)} and GLS periodograms of activity identifiers, to search for rotation period. \textbf{(b) -- (e):} H~\textalpha, Na~$D_{1}$, Na~$D_{2}$ and CCF FWHM. The plots use a logarithmic scale on the x-axes and focus on the region between 10 and 100 days. We expect the rotation period to lie in this range, based on initial stellar radius (Table~\ref{tab:dmpp-3aprops}) and rotation velocity values (Table~\ref{tab:rot_vels}). The shaded band highlights periods in the range 33 -- 36~d, where these indicators all share peak features.}
    \label{fig:rot_activity_GLS}
\end{figure}

We searched periodograms of the activity identifiers for signs of the rotation period between $15$ and $90$~d. The most significant periodicities were found in the analyses of H$\alpha$, Na~D1\,\&\,D2 and FWHM; as shown in Fig.~\ref{fig:rot_activity_GLS}. These indicators seem to share peaks at $\sim$\,33 -- 36~days. This might suggest that the rotation period lies in the range 33--36~d. A period of around 35~days would correspond to a $v\sin{i}$ of approximately 1.25~km\,s$^{-1}$. 
However, upon analysis of the spectral window function (Fig.~\ref{fig:rot_activity_GLS}\,a) for the same period range, we see that these activity periodograms look similar to the window plot. Especially for the FWHM, where there is not much periodic structure over shorter periods, the periodogram appears to be strongly affected by the sampling. When phase-folding the activity series on the identified periods, it becomes clear that sampling effects are at play. The points cluster into two distinct groups (with the majority of points in a single region), and phase coverage is very obviously uneven. These effects are characteristics of time series data folded on alias periods.

It is perhaps unwise to draw conclusions on the rotation period of the star from such activity analysis, when the temporal sampling is so influential. Further study is warranted, however a different approach might be needed: either through observations specifically scheduled to reduce aliasing; or a different method to determine the rotation period given that this is a slowly rotating, low activity star.

\section{Discussion}\label{sec:discussion}
DMPP-3 is an exciting discovery which offers several potentially productive avenues to explore. Most of these are phenomena we anticipated or could have anticipated when the Dispersed Matter Planet Project was conceived \citep{Haswell2020}, but the dynamically impossible 800\,d RV signal, Signal\,3, is a surprise. We begin our discussion by considering how it might arise. We did consider the possibility Signal 3 could be an instrumental artefact, but this possibility is quite easy to dismiss, given the excellent stability of {\sc HARPS}, the many other stars which have been monitored over a similar temporal baseline, and our analyses of the various observed line profile properties.

\subsection{Activity sources}\label{subsec:activitysources}
Stellar activity can affect RVs, even in relatively inactive old stars such as DMPP-3A. The main contribution to activity induced RVs for slowly-rotating, low activity stars is the suppression of convective blueshift \citep{2020A&A...Cretignier,2021MNRAS...Costes}. This is also the dominant effect in the Sun's RV variability \citep{2010A&A...512A..39..Meunier}. Convective blueshift (CB) is caused by up-welling of material in a star's convection granules on the photosphere. The rising of hot (bright) material seen at the centre of the region outweighs the effect from the cooler falling material, resulting in a net blue-shifted effect in the spectral lines. In active regions, magnetic fields hinder the convection and suppress the rising of new material, causing stellar regions such as spots and plages to be red-shifted in comparison to the surrounding disc \citep{2018A&A...Bauer}. The mean line profile used to determine the RVs for the star will therefore include shifted lines from the individual regions of suppressed CB, introducing an overall broadening (that will affect the RVs), that will be reflected in an altered FWHM measurement. \citet{2020A&A...Cretignier} simulate stellar activity for slowly rotating stars and find that the changes in RVs caused by diminished CB inside faculae likely dominates over the RV changes caused by the reduced flux of dark spots.

Convective blueshifts of stars with HARPS have been calculated recently by \citet{2021..Florian..CB}, who find empirical relations for the scale factor between CB on the Sun and other stellar types. For K0 type stars, like DMPP-3A, the relation predicts a value of $0.409$~CB$_{\sun}$. 

Spectroscopic and photometric variations of stars with periodicities $<100$~d will primarily be induced by the movement of active regions as the star rotates, and hence show modulation on the timescale of the rotation period. 
For longer periods (longer than the evolution timescale of individual activity features) the measured activity change would correspond to the activity level changing in a similar way to the solar magnetic cycle modulation on the Sun. The overall convective blueshift is dependent on the global activity level \citep{2017A&A...597A..52Meunier}. Therefore if the activity level changes over the course of a stellar cycle, the convective blueshift should roughly trace the global activity level. The amplitude of variation from CB over the Sun's magnetic cycle is 11~m~s$^{-1}$ \citep{2010A&A...512A..39..Meunier}, and by comparison the full amplitude of Signal 3 is $\sim$7~m~s$^{-1}$ (Table~\ref{tab:objects_params}). This value is broadly consistent with the relation from \citet{2021..Florian..CB} discussed above, where the CB effect will be reduced compared to the effect seen on the Sun. Therefore, whilst the $\sim$\,800\,d Signal 3 cannot arise from rotation, it perhaps may arise from activity modulation in DMPP-3A.

\subsection{Stellar activity cycles}\label{sec:cycles}
A modulation period of $\sim800$~d ($\sim2.2$~yr) is much shorter than
the Sun's magnetic cycle (approximately 11 yr). However, the cycle period changes from star to star, depending on numerous factors such as spectral type and binarity. A recent investigation into stellar cycles discusses stars with identified cycle periods just over 1000~d \citep{2022MNRAS.514.2259Sairam}. As our Signal 3 seems too long for rotation, it is likely to be a short stellar cycle. We can compare with \citet*{2016A&A...595A..12S-M}, who studied magnetic cycles of 12 K-type stars, and found a mean cycle length of 6.7~years. Two cycles identified in stellar photometry were however 1.7 and 2.7~years, similar in length to our 2.2~year Signal 3.

\citet{2016A&A...595A..12S-M} also identify some stars with multiple cycles. For example, GJ\,729 has two superimposed cycles, of 7.1 and 2.1~years. The authors suggest that the shorter period might be a "flip-flop" cycle. A
flip-flop cycle is the repeated inversion in longitude of spot regions on the stellar photosphere. This effect was first studied on the Sun, where spot activity alternates in longitude on timescales of 1.5 to 3~years, causing cycles of 3.8 and 3.65~yr in the northern and southern hemispheres, respectively \citep{2003A&A...405.1121B}. These periods are roughly 1/3 of the 11~year sunspot cycle, a ratio that is seen to persist over long timescales. This flip-flop effect can be observed in both spectroscopic and photometric measurements \citep{2016A&A...595A..12S-M}.

The cycle observed on DMPP-3A with period of 2.21~years has a range of potential origins. The cause could be: a short stellar magnetic activity cycle; a sub-cycle of a longer cycle that we are not sensitive to detect; a harmonic of a longer cycle; or a flip-flop cycle. The cycle length is consistent with a flip-flop cycle 1/3 the length of a longer stellar magnetic activity cycle, given that the average found by \citet{2016A&A...595A..12S-M} for K-type stars was 6.7~years -- although this could be purely coincidental.

\subsection{Implications for RV detections}\label{sec:discuss_rvdet}
The lack of any well-defined periodicity at 800\,d in the S-index and the BIS is consistent with a dynamical interpretation of the 800\,d period. The absence of any systematic variation of the line bisector shapes on the 800\,d period is also a positive sign for a dynamical interpretation. We did see power in the S-index between 540\,d and 1000\,d and more localised and higher significance peaks around 800\,d in the FWHM and CCF area periodograms. It is possible these latter periodicities would have caused us to doubt the 800\,d signal was due to a planet even if it had been found in a single-star system. It is also possible that we would have attributed Signal 3 to a planet after examining the S-index and BIS periodograms and the line bisector shapes. It was the impossibility of the dynamical explanation of Signal 3 which motivated a very detailed examination of all the available activity indicators, along with a consideration of the long-term effects observed in the Sun.

This raises the possibility that there may be published planet `detections' which have an origin analogous to that of Signal 3. The techniques for discerning the long-period spectroscopic signatures of stellar activity cycles are only now being developed. Signal 3 in DMPP-3AB provides a cautionary example and an opportunity to learn. Further monitoring of DMPP-3 may reveal more clear-cut evidence that Signal 3 is not strictly periodic, as the S21 data already begins to suggest (cf. Fig.~\ref{fig:sBGLS}). If we are to confidently detect long period planets, including analogues of the Solar System planets, we must hope that this is the case. This is perhaps our most important conclusion as it has serious and widely-applicable implications.

\subsection{Consequences of a second planet}\label{sec:discuss_2ndplanet}

If the 2.26\,d period Signal 4 is a planet, DMPP-3 becomes an even more exciting prospect for study into planetary system formation, evolution and dynamics.
The tightest S-type planetary system currently known to contain multiple planets\footnote{Planets in binary systems are recorded in a machine-readable table, accessed from~\url{https://lesia.obspm.fr/perso/philippe-thebault/plan_bin500au.txt}. The table is complete for all planets on S-type orbits in binaries of separations up to 500~au (last updated 2022 Sept 01, contains 171 systems).} (strictly, a gas giant and a brown dwarf) is HD\,87646, where the primary star in a binary of separation 19.5~au hosts two substellar objects \citep{2016..Ma..HD87646}. A second planet in DMPP-3 would confront and challenge ideas about planetary system evolution. A major factor in governing evolution in binary star planetary systems is the separation; the nearest multi-planet analogue has a binary separation over an order of magnitude higher than that of DMPP-3.

The two close-in S-type planets will interact gravitationally with each other as well as with the very low mass stellar eccentric binary companion DMPP-3B. Their orbits will evolve under dynamic excitation of eccentricity and tidal circularisation. They potentially offer opportunities to derive constraints on the material properties of the objects, including the viscoelasticity and tidal response. This ultimately could offer some empirical clues to the interior structure of the two low mass planets, cf. \citet{Makarov2018}. This is particularly interesting as these planets are likely to be losing mass, and may be the remnant cores of previously more massive objects. They lie below the Neptune desert in the orbital period -- planet mass plane. Furthermore, transmission spectroscopy may reveal the chemical composition of the vaporising planetary surface, complementing the information deduced from the tidal properties. 

Our potential detection of a second S-type planet highlights the power of the DMPP target selection.  A mere ten years ago, \citet{2012A&A...542A..92Roell} made conclusions on the state of discoveries of exoplanets in multiple stellar systems, with two of the main points being that so far planets hadn't been detected in close binaries with separation $< 10$~au, and that multiple planets had not been detected in systems with separation $< 100$~au. DMPP has identified a system which defies these limitations thus providing an excellent laboratory to study extra-solar system formation and evolution.

\subsection{\textbf{Planet formation in a tight binary}}
The evolutionary history of the DMPP-3 system is an interesting topic for debate.
Much consideration has been given to the circularisation of eccentric binary star orbits by tidal forces, with a comprehensive recent summary of the theoretical work given in the introduction of \citet{2022ApJ...929L..27Z}.
Tidal forces are obviously most influential for short orbital periods, and observational selection effects make systems with short orbital periods more likely to be discovered and further studied. \citet{2022ApJ...929L..27Z} accordingly do not consider periods as long as that of DMPP-3AB. \citet{2022MNRAS.tmp.3049C} use a comprehensive sample of eclipsing binaries to conclude that long period binaries may be tidally circularised significantly more efficiently than is usually assumed. DMPP-3AB is more eccentric than any of their sample  - cf. fig.~6 of \citet{2022MNRAS.tmp.3049C} - but offsetting the high eccentricity, DMPP-3B has the lowest mass possible for a star. It is therefore difficult to rule out the stellar binary existing in more or less the current configuration for the entire main sequence lifetime of the primary star. HD\,137496 is orbitally similar to DMPP-3, with a dense, hot super-Mercury on a 1.6\,d orbit and a cold $M_p\sin{i} \approx 7.7 \, \rm{{M}_{J}}$ giant on a 480\,d orbit with $e = 0.477 \pm 0.004$ \citep{2022A&A...657A..68A}. It is possible these two systems share aspects of their evolutionary history.

\subsubsection{In situ planet formation}\label{subsub:insit}
Assuming the system began in essentially the present configuration, we can consider the likelihood of in-situ planet formation. Though they orbit in the semi-empirical `safe' zone for S-type planets \citep{Holman1999}, the available mass for DMPP-3A\,b (Signal 2) and the putative Signal 4 planet to form out of would have  been limited by truncation of the protoplanetary disc. Theoretical models of tidal truncations (for dust discs in the Taurus region) are described by \citet{2019Manara}, who provide an analytical function for truncation radius depending on input binary parameters. Using their equation~C.1 for the DMPP-3 system, we are able to determine truncation radii for differing Reynolds numbers (which inform the analytically derived coefficients used for a $\mu\sim0.1$ binary). The Reynolds number is related to the magnitude of viscous stress, which resonant torques need to overcome in order to truncate the disc \citep{Zeng2022}.

The resulting radii are $<0.4$~au, for Reynolds numbers $10^{4}$--$10^{6}$. \citet{Zeng2022} show in their study of the Gliese-86 system that truncation radius decreases with increasing Reynolds number, and consider Reynolds number in the range in  $10^{3}$--$10^{14}$. We thus take a lenient upper limit of 0.4~au as the truncation radius.
To form the $\sim2$~M$_{\earth}$ planet from a disc with radius $0.4$~au, we would require a mean dust surface density of $\Sigma_{\textrm{d}} \sim100$~g~cm$^{-2}$. \citet{Tazzari2017} investigate dust densities in protoplanetary discs (in the Lupus star forming complex) observed with ALMA. Through use of their equations~1\&2 and data in their table~3 we can determine mean $\Sigma_{\textrm{d}}$. For the 22 systems included in \citet{Tazzari2017}, we calculate $\Sigma_{\textrm{d}} \lesssim2$~g~cm$^{-2}$.
For the discs studied, there seems to be no relationship between disc parameters and stellar parameters \citep{Tazzari2017}. We can conclude that unless the disc that formed DMPP-3A\,b was anomalously dense, it is therefore unlikely that the system formed in the current configuration. This conclusion is reinforced by the likelihood that the hot planet(s) may also have been losing mass since their formation.

\subsubsection{Mass loss}\label{subsub:massloss}
DMPP target stars were selected through the spectroscopic signature of circumstellar gas attributed to mass losing hot planets \citep{Haswell2020}. DMPP-3 is $\sim$10 Gyr old; the planet(s) would need to initially be more massive if they have been continually losing mass. CDEs provide the most dramatic examples of mass loss, exemplified by Kepler-1520b. The planet is a hot, $\sim$\,$0.1\, {\rm M_{\earth}}$, rocky planet heated to $\sim$\,$2100$\,K, where extreme irradiation vapourises the rocky surface \citep{2012ApJ...752....1Rappaport}. Dust condenses from the metal-rich vapour and subsequently forms  a comet-like tail which causes variable-depth transits. The planet DMPP-3A\,b has an estimated equilibrium temperature of $T_{\rm eq} \sim 850$~K \citepalias{Barnes2020}. The putative Signal 4 planet would have $T_{\rm eq} \sim 1800$~K, almost as hot as Kepler-1520\,b.

Temperature is however not the only factor that dictates the mass-loss rate for a disintegrating planet. \citet{2013MNRAS.433.2294P}  model the mass-loss history of a CDE for different initial masses. The larger the formation mass of the planet, the more likely it is to retain material due to the stronger surface gravity. \citet{2013MNRAS.433.2294P} found for $T=2145K$, rocky planets of initial mass above $\sim0.12~{\rm M_{\earth}}$ are able to survive for longer than $10$~Gyr (see their figure~9). 

\citet{Booth2022} extend the simulations of CDE mass-loss history. For the first time, they include models for the formation of the dust grains, as well as progressing the treatment of dust heating by considering both stellar and re-emitted thermal radiation from the planet itself. Their analysis builds on previous work, and predicts mass-loss rate for a range of planet masses \textit{and} temperatures (Figure~6; \citealt{Booth2022}). This then implies how long the planets would survive under gas and dust loss. The shortest lived planets are the hottest and smallest, in agreement with previous works. 

The planet(s) in the DMPP-3 system, despite being heated to high temperatures, have minimum masses that are far above the threshold required for significant mass loss. The putative Signal 4 planet has mass $M_{\textrm{p}}>0.8~{\rm M_{\earth}}$, in comparison with the far lower formation mass of $\sim0.05~{\rm M_{\earth}}$ required for it to not survive 10~Gyr at 1800~K (cf. Fig.~6 in \citealt{Booth2022}). The mass-loss would be negligible in comparison to the total mass of the planet, and the age of the system. Therefore, solely from an evaporation history viewpoint, it would be plausible that the planet(s) could reside in the current orbit(s) whilst retaining the majority of their material.

\subsubsection{Circumbinary capture}\label{subsub:p-type-capture}
The protoplanetary disk truncation arguments (Section~\ref{subsub:insit}) imply that DMPP-3 has undergone dynamical reconfiguration. Formation of planets outside a close binary would be much easier to accomplish, and is feasible for distances far enough away from the central stars \citep{Meschiari2012}. There exists an instability zone outside the binary where the companion stirs up eccentricity of planetesimals causing higher encounter velocities. Faster impacts reduce the likelihood of forming large bodies through accretion \citep{Paardekooper2012}. \citet{Holman1999} simulated the critical semi-major axis around the inner binary where a planet will always be stable, and using their semi-empirical formula we find a P-type critical semi-major axis of $3.95$~au for the DMPP-3 system.

Some circumbinary planets have been found orbiting within this `critical semi-major axis' in their system, e.g. Kepler-16\,b \citep{Meschiari2012}. The stability limit for a system also depends on mean motion resonances, with regions of stability in-between first-degree ($N:1$) resonances \citep{2018ApJ..Quarles,2022MNRAS.512..602M}. The most likely evolutionary scenario would be the formation of a core far enough away from any unstable regions, followed by migration inwards (often quickly passing through unstable resonances) due to gravitational interactions -- with either the protoplanetary disk or a second circumbinary planet \citep*{Meschiari2012, 2022MNRAS.512.5023F}.

Through migration towards the centre of the system, there is a possibility that such a planet could then be captured by one of the components of the binary. This was investigated by \citet{2018MNRAS...Gong}, who found that there was a low (but non-zero) probability of such a scenario occurring. The probability is improved with planet-planet scattering before a capture. This process would require an instability mechanism to stir up chaotic motions for any chance of capture -- such as destabilising mean motion resonances (e.g. 5:1, 6:1, 7:1) with the inner binary \citep{2022MNRAS.512..602M}, or mutual inward resonant migration of two P-type planets forcing an inner planet too close to the central stars \citep{2022MNRAS.512.5023F}. A P-type to S-type conversion would be very rare, but could be possible provided that the unstable motion does not end with ejection or collision with one of the stars \citep{2016ApJ...818....6S}.

P-type planet capture seems plausible, albeit unlikely, for a single planet, but not so for two planets in an S-type orbit around the primary. The likelihood for two planets to be captured, and not ejected during any interactions, will then be far smaller. Further RV observations of DMPP-3 to confirm the putative second planet are needed to reveal whether or not this formation channel is viable.

\subsubsection{Alternative initial configurations}
We can also consider migrations within the system. The current location of the binary and the instability zone it creates \citep{Holman1999} would mean that for any planetary migrations from wider orbits, the binary star would also have needed to start life on a wider orbit. Interactions with protoplanetary discs can cause migrations that allow formation of planet pairs in mean motion resonance (MMR) despite the presence of a binary companion. \citet{2022..Roisin} simulate the evolution of planet pairs, and find that resonances (such as the 3:1 resonance, close to the orbital period ratio between Signal 2 to Signal 4) can arise during migrations. 

If DMPP-3B formed further out and migrated inwards, perhaps caused by interaction with an external object, then it would provide a safer environment for the circumprimary planet(s) to form. Low mass hydrogen burning secondary stars are typically formed through fragmentation of protostellar accretion disks \citep{2012ApSS.341..395..Kaplan}. This fragmentation tends to happen on scales of $\sim 100$~au, but a lot of these very low mass stars (VLMSs) end up close to the primary through scattering and secular migration \citep{2012ApSS.341..395..Kaplan}.

The final scenario to consider here in the formation and evolution discussion is the possibility that the binary companion was not present during the formation of the planets. It seems plausible for planets to form (and potentially migrate inwards) close enough to the parent star, where they could remain safe during a capture of a more massive body at some point during the system's history. Exchange reactions with another system could potentially swap existing companions in four-body interactions, but would be rather unlikely \citep{2012ApSS.341..395..Kaplan}. Tidal capture could occur between two objects, creating a binary system, which would tend to produce tight binaries \citep{2011psf..book.....B}. A chaotic capture of a VLMS by a Sun-like star could also be the mechanism that provided the large eccentricity we observe for DMPP-3B.

\section{Conclusions}\label{sec:conclusions}

We have studied the dynamics of the compact, eccentric S-type binary DMPP-3. New observations allowed us to study the reflex radial velocities and examine the signatures of stellar activity in more depth than previous work. Our main conclusions are:

\begin{enumerate}
    \item We derive significantly more precise parameters for the DMPP-3AB binary orbit. The 3$\sigma$ lower limit on the projected mass of DMPP-3B is now 80.9\,M$_{\textrm{jup}}$. This establishes DMPP-3B can sustain hydrogen burning, and is a star at the very bottom of the main sequence. 
    
    \item The dynamically problematic 800\,d RV signal identified by \citetalias{Barnes2020} is confirmed, though the stacked periodogram suggests the signal may not be strictly periodic. 
    
    \item Numerical simulations demonstrate that there is no mutual inclination for which DMPP-3AB can harbour an object producing the 800~d signal via reflex radial velocities. This confirms and strengthens the conclusion of \citetalias{Barnes2020}, that the 800\,d signal must arise from stellar activity.
    
    \item Comprehensive investigation of the activity indicators provides evidence (from the S-index, FWHM, and CCF area) that the $\sim$800\,d Signal 3 is an artefact of stellar activity. 

    \item We confirm the detection of the 6.67\,d S-type super-Earth planet DMPP-3A\,b and refine its parameters, finding 
    $M_{\rm p}\sin{i} =2.224^{+0.502}_{-0.279} {\rm M}_{\earth} $.
    
    \item An additional 2.26\,d Earth mass S-type planet candidate is tentatively detected (Signal 4; 0.2 per cent FAP; 2$\sigma$ significance). Being both hotter and lower mass than DMPP-3A\,b this planet candidate would be more likely to produce radiation-driven mass loss, and create a diffuse circumstellar gas shroud. Further high precision RV observations are required to confirm this planet candidate.
    
    \item There is no sign of either DMPP-3A\,b or Signal 4 being due to stellar activity, and orbital simulations demonstrate stability for a two-planet system.
    
    \item The DMPP-3AB binary is detected astrometrically by \textit{Gaia}. The resulting orbital inclination, $63.89 \pm 0.78 ^{\circ}$, allows us to constrain the mass of DMPP-3B to $91.90 \pm 0.85$~M$_{\textrm{jup}}$. If the planet(s) lie in the same orbital plane, we can estimate `coplanar' masses of $M_{\rm A\,b} = 2.47 \pm 0.56$~M$_{\earth}$ and $M_{\rm Sig.4} = 1.186 \pm 0.289$~M$_{\earth}$
    
    \item There may be published long period planet `detections' which have an origin analogous to that of Signal 3. Further RV monitoring of DMPP-3 will reveal signatures which can be used to most efficiently identify these imposters. This is perhaps our most important conclusion as it has serious and widely-applicable implications.
    
\end{enumerate}

\section*{Acknowledgements}
These results were based on observations made with the ESO 3.6~m telescope and HARPS, under ESO programme IDs: 081.C-0148(A); 088.C-0662(A); 091.C-0866(C); 096.C-0876(A); 098.C-0269(A); 098.C0499(A); 098.C0269(B); 099.C-0798(A); 0100.C-0836(A). Specifically, the new observations reported here were obtained under ESO programmme 107.22UN. These observations were obtained in service mode by staff at ESO La Silla Observatory. 

The authors thank Matthew R. Standing and the anonymous referee for useful comments which greatly improved the quality of this manuscript. ATS and ZOBR are supported by STFC studentships. CAH and JRB are supported by grants ST/T000295/1 and ST/X001164/1 from STFC. JKB is supported by an STFC Ernest Rutherford Fellowship (grant ST/T004479/1). This research has made use of the SIMBAD data base, operated at CDS, Strasbourg, France. Simulations in this paper made use of the \textsc{rebound} code which can be downloaded freely at \url{http://github.com/hannorein/rebound}. Stellar parameter estimation in this work has made use of the \textsc{species} software (\url{https://github.com/msotov/SPECIES}). Radial velocity data were analysed with the \textsc{exo-striker} software (\url{https://github.com/3fon3fonov/exostriker}). 

This work has made use of data from the European Space Agency (ESA) mission {\it Gaia} (\url{https://www.cosmos.esa.int/gaia}), processed by the {\it Gaia} Data Processing and Analysis Consortium (DPAC, \url{https://www.cosmos.esa.int/web/gaia/dpac/consortium}). Funding for the DPAC has been provided by national institutions, in particular the institutions participating in the {\it Gaia} Multilateral Agreement.

\section*{Data Availability}


The data underlying this article are available in Open Research Data Online (ORDO; \url{https://ordo.open.ac.uk/}), at \url{https://doi.org/10.21954/ou.rd.21324549.v1}. The datasets were derived from HARPS spectra in the public domain, accessed from ESO Phase 3 Archive (\url{http://archive.eso.org/wdb/wdb/adp/phase3_main/form?phase3_collection=HARPS}).


\bibliographystyle{mnras}
\bibliography{references}



\appendix

\section{Orbital Configuration}\label{appendix:orbit}

\begin{figure*}
	\includegraphics[width=0.48\linewidth]{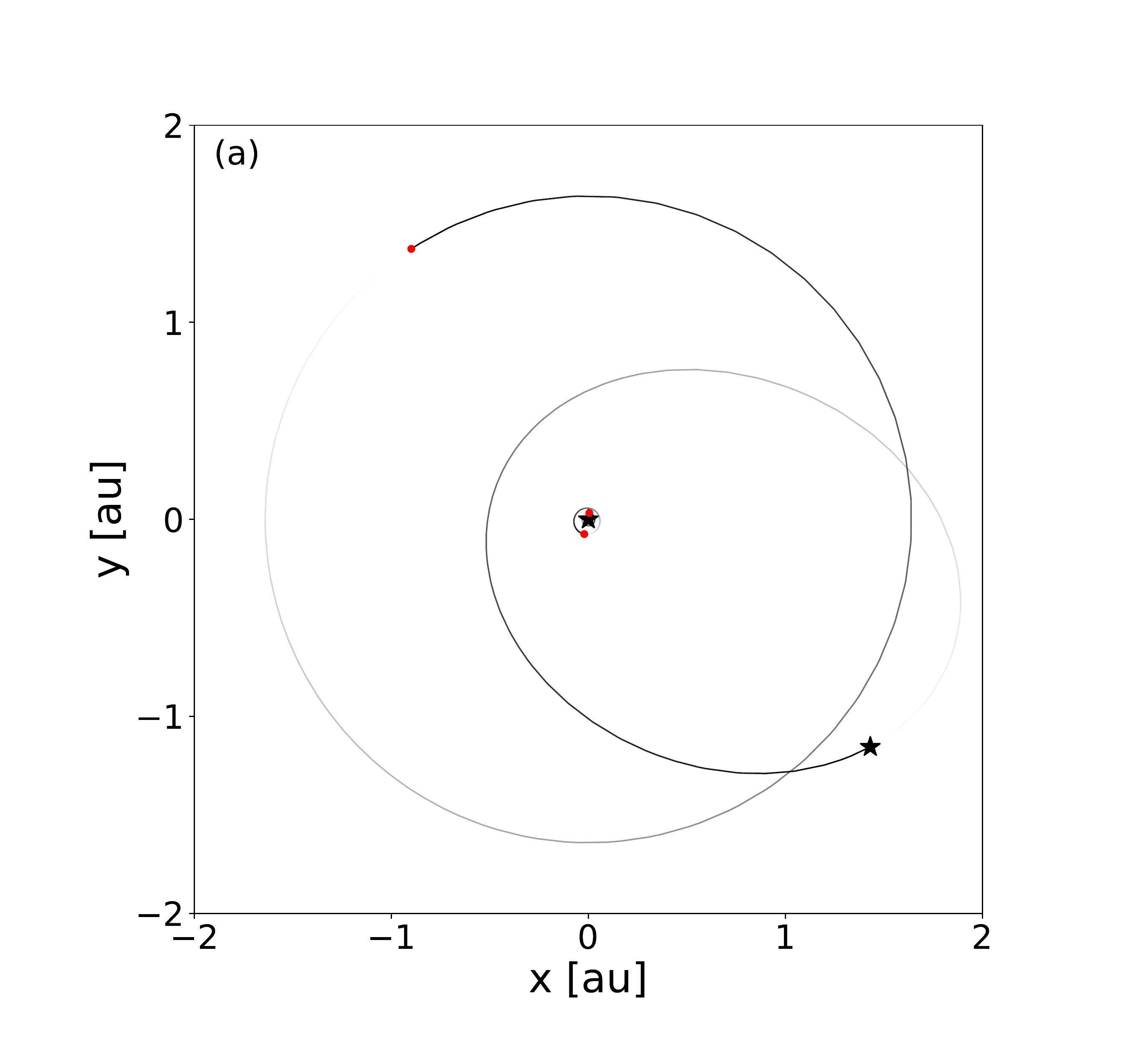}
	\includegraphics[width=0.48\linewidth]{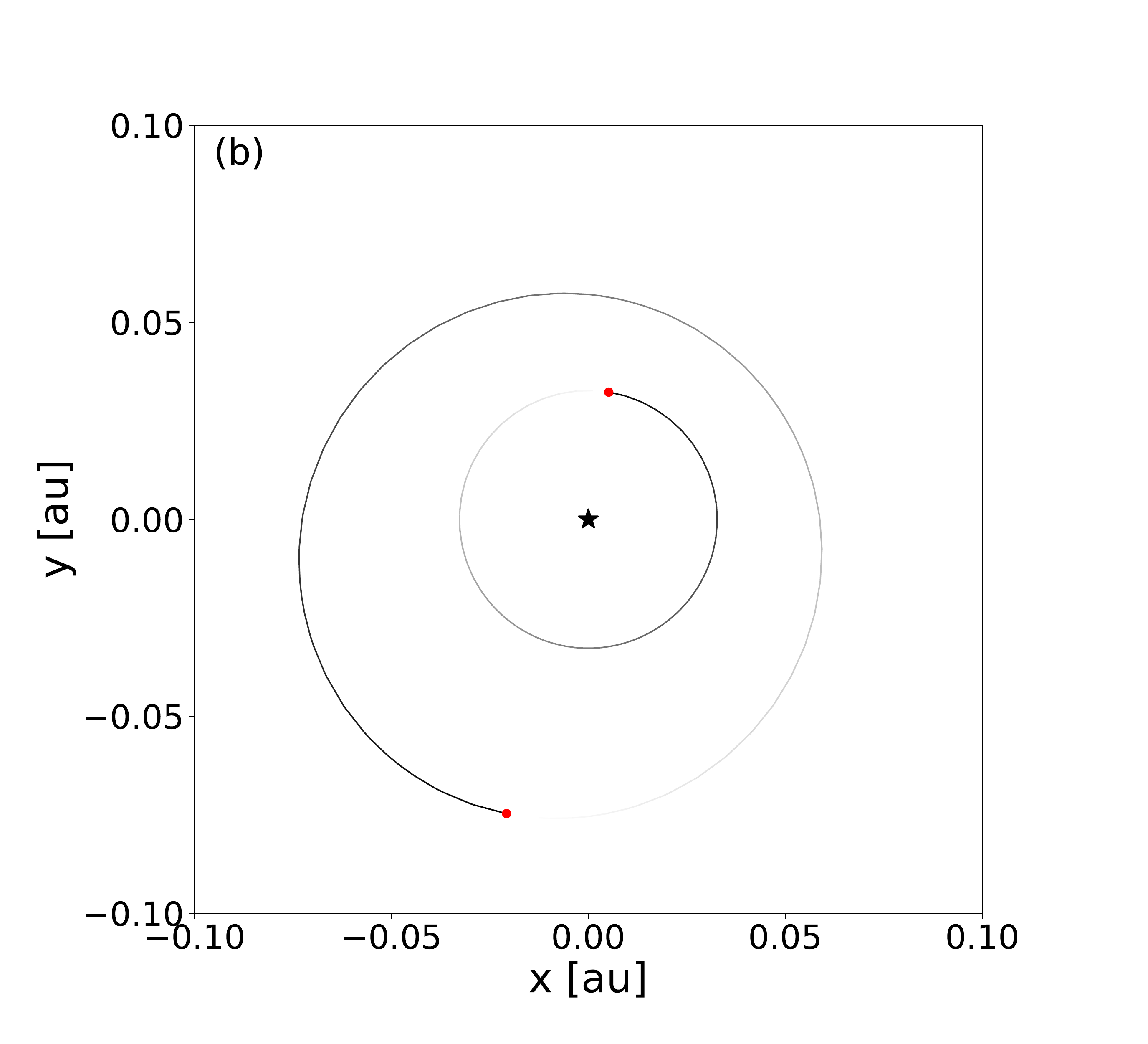}
	\caption{Diagrams showing the DMPP-3 orbital configuration, if Signal~3 is indeed an orbiting body. Stellar objects shown as black star symbols, and planets as red circles. Objects not shown to scale relative to orbit sizes. The axes of $x$ and $y$ help visualise the sizes of the orbits, but correspond to arbitrary directions in the orbital plane \textbf{(a)} The entire view. \textbf{(b)} An enlarged section near the star, showing the possible orbits of two S-type planets.}
    \label{fig:orbits}
\end{figure*}

To visualise the orbital configuration of the DMPP-3 system, diagrams detailing this are shown here in Fig.~\ref{fig:orbits}. The entire system is shown, as is the zoomed-in section near the star where the planet(s) exist in stable orbits after orbital evolution simulations. The orbits are plotted with parameters corresponding to the maximum \textit{a posteriori} parameters, as described in Table~\ref{tab:objects_params}. The orbits are shown for the co-planar case, and it is immediately apparent that if Signal 3 was attributed to a planet there would be a collision with the binary companion. Since orbital simulations (Section~\ref{sec:3Dsims}) show that for any mutual inclination of companion and `800~d object' the system is unstable, this is only included to illustrate the scenario.

\section{FWHM corrections}\label{appendix:fwhm}
\begin{figure*}
	\includegraphics[width=0.90\linewidth]{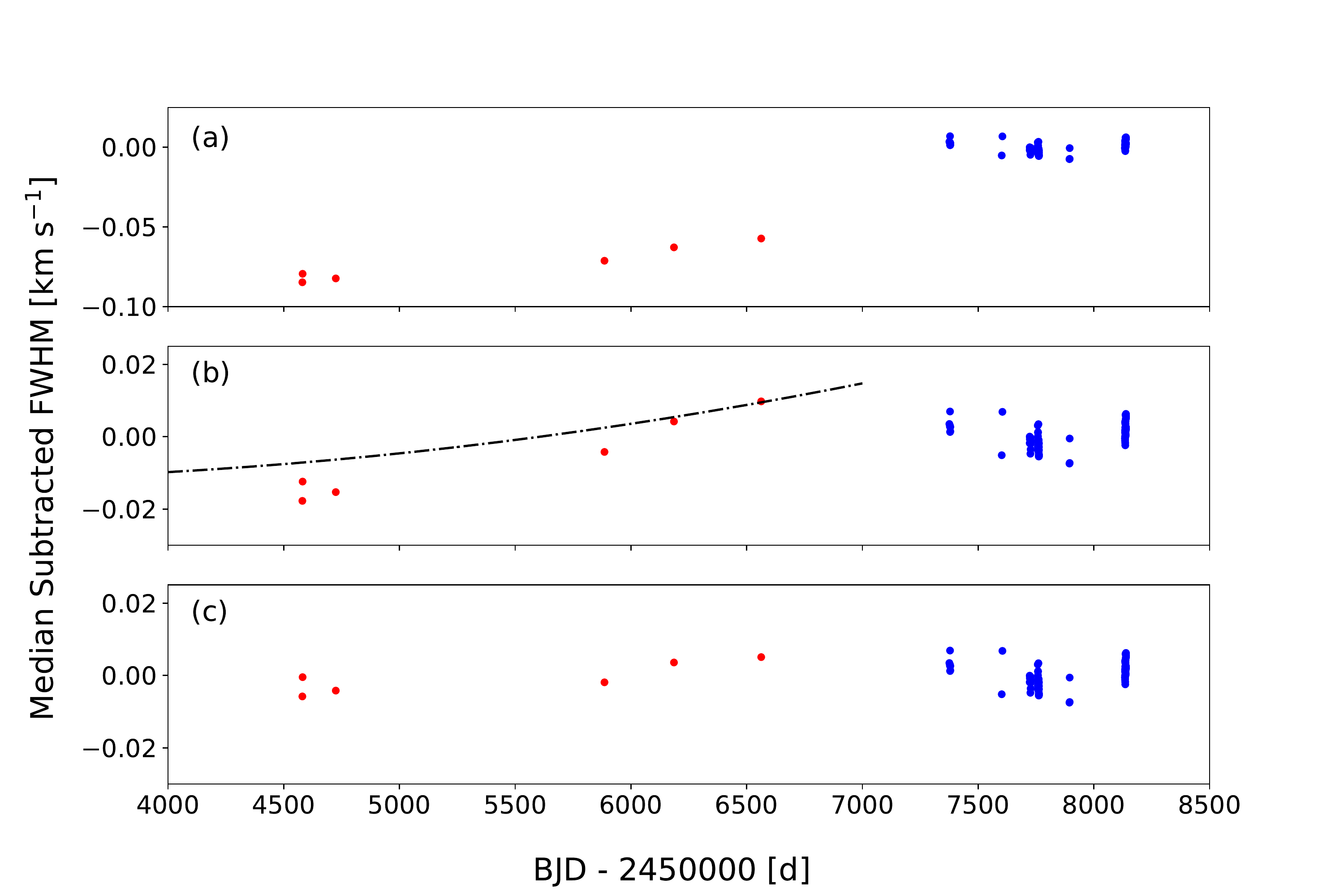}
    \caption{Visualisation of the corrections performed on the FWHM data before and after the fibre upgrade (at BJD $\sim$ 2457174). The 6 points on the left (red) are the pre-upgrade FWHM values. \textbf{(a)} The uncorrected values, focusing on the CHEPS and DMPP datasets. \textbf{(b)} Normalising before and after upgrade to a common zero median, treating as two independent datasets. The trend from \citet{2021MNRAS...Costes} is plotted over in dashed lines, and extends until the era of the fibre upgrade. As can be seen from the main DMPP observations taken after the upgrade, they do not appear to follow a similar polynomial trend. \textbf{(c)} The pre-upgrade values with the trend now subtracted off, bringing both datasets together coherently. }
    \label{fig:fwhmtrend}
\end{figure*}

We used eight archival measurements made with {\sc HARPS} between 2008 and 2013 \citep{2009MNRAS.398..911J..Calan} as well as many more recent observations. Thus the observations span the {\sc HARPS} fibre exchange in 2015 \citep{HarpsUpgrade}. We see a discontinuity in the FWHM (Fig.~\ref{fig:fwhmtrend}\,a). We normalised pre- and post-upgrade data by subtracting the median FWHM value of each (Fig.~\ref{fig:fwhmtrend}\,b). This corrects the apparent $0.07\, \textrm{km}\,\textrm{s}^{-1}$ offset but could, however, absorb some real astrophysical differences.

Despite this, we anticipated the correction would not significantly impact the search for periodicities on timescales less than 1000 days as there are only 8 pre-upgrade points spanning 5 years, two of which were ultimately discarded (see Appendix~\ref{appendix:outliers}).

\begin{figure*}
\centering
	\includegraphics[width=0.95\linewidth]{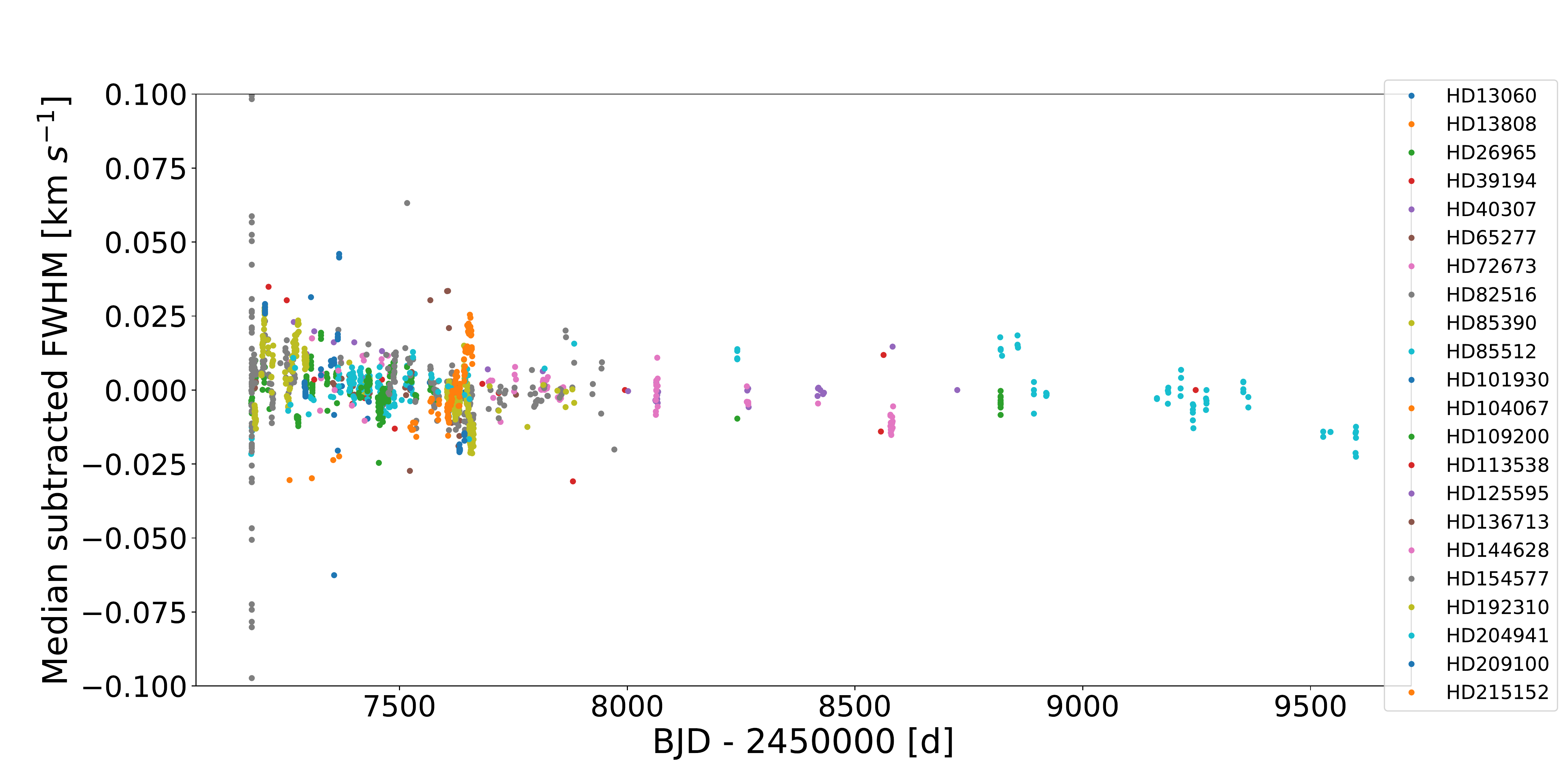}
    \caption{A plot of FWHM, showing the post-upgrade values for the K-type stars included in the \citet{2021MNRAS...Costes} study. Alpha Centauri B is not included here due to the vast data volume and hard-disk space needed to store the $\sim$ 18000 spectra on record. Each star's data is plotted in a different colour, and it can be seen that there is no obvious long term trend identified from the data available, in the manner that there was for pre-upgrade measurements in \citet{2021MNRAS...Costes}.}
    \label{fig:nomorefwhmtrendkstars}
\end{figure*}

The second correction to make is due to the systematic long-term drift in the focus of the instrument \citep{2018A&A...Dumusque}. In a recent analysis by \citet{2021MNRAS...Costes}, the authors analyse a sample of main sequence stars, where they identify the HARPS instrumental focus drift and apply a correction based on spectral class for FGK stars. A polynomial trend is computed for the stars, but their study only uses data points before the fibre upgrade. This is given, with empirically determined coefficients for K-type stars in Table~\ref{tab:trendcoef}, as:

\begin{equation}
    a\,\times\,(\textrm{BJD} - \textrm{BJD$_{0}$})^{2}\,+\,b\,\times\,(\textrm{BJD} - \textrm{BJD$_{0}$})\,+\, c. 
    \label{eqn:trend}
\end{equation}
\noindent Here $a$, $b$, and $c$ are said coefficients, BJD denotes the observation time, and BJD$_{0}$ the reference initial time (2452937.57).
\begin{table}
	\centering
	\caption{Coefficients for use in Equation~\ref{eqn:trend}. These were determined by \citet{2021MNRAS...Costes} through analysis of HARPS data for 23 K-type RV standard stars, all with a high number of spectra recorded over a long baseline.}
	\label{tab:trendcoef}
	\begin{tabular}{lc} 
		\hline
		Variable & Value ($\textrm{km}\,\textrm{s}^{-1}$) \\
		\hline
		a  & $1.5 \pm 0.2 \times 10^{-9}$  \\
		b & $4.9 \pm 8.9 \times 10^{-7}$  \\
		c  & $-1.2 \pm 0.1 \times 10^{-2}$  \\
		\hline
	\end{tabular}
\end{table}
%
%
%
%
We used this polynomial to remove the trend from the CHEPS data points (Fig.~\ref{fig:fwhmtrend}\,c). To extend the analysis of \citet{2021MNRAS...Costes} we used the same sample of stars to determine that the observed trend does not continue after the fibre upgrade (Fig.~\ref{fig:nomorefwhmtrendkstars}). We already suspected this from the DMPP-3 post-upgrade FWHM data, and we performed no further corrections to the FWHM data.

The FWHM in the analysis presented above and in Section~\ref{sec:fwhmanal} comes directly from the {\sc HARPS} {\sc drs}. We tried also performing a colour correction, using the flux templates from the {\sc HARPS} reduction process to re-weight the CCF orders (A. Suarez-Mascareño, private communication). A  re-weighting of the CCF from the {\sc drs} improves results for M-dwarfs \citep{2021A&A...648A..20Toledo}. For our data this alternate approach did not provide improvement upon that from {\sc drs}, which was to be expected as the {\sc HARPS} {\sc drs} already implements a colour correction  for all except M-type stars \citep{2014SPIE.9147E..8CC}.

\section{Two Outliers}\label{appendix:outliers}
A target radial velocity needs to be specified for the {\sc drs} pipeline to successfully extract RV values. If the target RV field is set to 99999 (km~s$^{-1}$), RV calculation can be turned off, and the data products are calculated incorrectly. Two of the CHEPS spectra are flagged in this way and have outlying FWHM, BIS and contrast values. We have rejected them from these activity time-series. Note: the RVs themselves were derived using the \textsc{terra} software, and we have no reason to exclude them. 


\bsp	
\label{lastpage}
\end{document}